\documentclass[twocolumn]{aastex63}
\usepackage[normalem]{ulem}  
\usepackage[T1]{fontenc}
\usepackage{ae,aecompl}
\usepackage{graphicx}
\usepackage{amsmath}
\usepackage{amssymb}
\usepackage{supertabular}
\usepackage{hyperref}
\usepackage{xcolor}
\usepackage{academicons}
\usepackage{multirow}

\def\Msun{\,M_{\odot}}
\def\mev{\;\text{MeV}}
\def\fm3{\;\text{fm}^{-3}}
\def\mev{\;\text{MeV}}

\begin{document}

\title{Astrophysical implications on hyperon couplings and hyperon star properties with relativistic equations of states}


\author{Xiangdong Sun}
\affiliation{Department of Astronomy, Xiamen University, Xiamen, Fujian 361005, China; liang@xmu.edu.cn}

\author[0000-0003-1197-3329]{Zhiqiang Miao}
\affiliation{Department of Astronomy, Xiamen University, Xiamen, Fujian 361005, China; liang@xmu.edu.cn}

\author[0000-0001-8958-9787]{Baoyuan Sun}
\affiliation{Frontiers Science Center for Rare Isotopes, Lanzhou University, Lanzhou 730000, China}

\author[0000-0001-9849-3656]{Ang Li}
\affiliation{Department of Astronomy, Xiamen University, Xiamen, Fujian 361005, China; liang@xmu.edu.cn}


\begin{abstract}
Hyperons are essential constituents in the neutron star interior. The poorly-known hyperonic interaction is a source of uncertainty for studying laboratory hypernuclei and neutron star observations. In this work, we perform Bayesian inference of phenomenological hyperon-nucleon interactions using the tidal-deformability measurement of the GW170817 binary neutron star merger as detected by LIGO/Virgo and the mass-radius measurements of PSR J0030+0541 and PSR J0740+6620 as detected by NICER. The analysis is based on a set of stiff relativistic neutron-star-matter equation of states with hyperons from the relativistic mean-field theory, naturally fulfilling the causality requirement and empirical nuclear matter properties.
We specifically utilize the strong correlation recently deduced between the scalar and vector meson hyperon couplings, imposed by the measured $\Lambda$ separation energy in single-$\Lambda$ hypernuclei, and perform four different tests with or without the strong correlation. We find that the laboratory hypernuclear constraint ensures a large enough $\Lambda$-scalar-meson coupling to match the large vector coupling in 
hyperon star matter. When adopting the current most probable intervals of hyperon couplings from the joint analysis of laboratory and astrophysical data, we find the maximum mass of hyperon stars is at most $2.176^{+0.085}_{-0.202}M_{\odot}$ ($68\%$ credible interval) from the chosen set of stiff equation of states. The reduction of the stellar radius due to hyperons is quantified based on our analysis and various hyperon star properties are provided. \\
\end{abstract}

\section{Introduction}

It is well known that neutron stars are not made of pure neutrons, since some of them will beta-decay until equilibrium between neutrons $n$, protons $p$ and leptons (electrons $e$ and muons $\mu$) is reached, namely $\mu_p=\mu_n-\mu_e$, $\mu_n$, $\mu_p$ and $\mu_e$ being the chemical potential of neutrons, protons and electrons, respectively. 
Such uniform $npe\mu$ matter is the minimal model of the inner neutron-star core. 
Because the chemical potentials grow with increasing baryon density, additional particle species besides nucleons ($N$) will be populated when their thresholds are reached.
For example, $\Lambda$ hyperons, the lightest baryon octet ${\rm J}^{\rm P}=(1/2)^+$, may appear in high-density neutron-star cores via weak interaction process $p+e^-\longrightarrow\Lambda+\nu_e$, replacing highly energetic neutrons when the $\Lambda$ chemical potential fulfills
the condition $\mu_{\Lambda}=\mu_n=\mu_p+\mu_e$.
Other higher-mass hyperon species ($\Sigma$ and $\Xi$ hyperons) may also appear.
Although unstable under terrestrial conditions, hyperons (Y) are stable in dense stellar matter because the Pauli principle blocks their decay into nucleons. Consequently, the core of a massive neutron star consists of an inner hyperon core and an outer nucleon shell. These neutron stars are unusually called hyperon stars.

Ever since it is pointed out that hyperons could be present in neutron star cores in 1959~\citep{1959ApJ...130..916C}, enormous efforts have been done to learn the composition and equation of state (EOS) of neutron stars including hyperons. 
It is a great challenge since the results are very sensitive to the poorly-known hyperonic interaction and thus very model-dependent.
There are attempts to overcome the problem by exploiting arguments of strong-interaction symmetry. For example, in the widely-used relativistic field theoretic approaches, the hyperon-vector-meson coupling constants are unusually fixed from the SU(6) spin-flavor model in the forms of 
$g_{\omega N}:g_{\omega \Lambda}:g_{\omega \Sigma}:g_{\omega \Xi}=3:2:2:1$ and $g_{\rho N}:g_{\rho \Lambda}:g_{\rho \Sigma}:g_{\rho \Xi}=1/2:0:1:1/2$, $\omega$ and $\rho$ being the isoscalar–vector and isovector–vector mesons, respectively.
Then $g_{\omega Y}$ and $g_{\rho Y}$ are determined once $g_{\omega N}$ and $g_{\rho N}$ are known.
Experimentally, hyperonic interaction could be derived from the scattering experiments (several hundred $\Lambda N$ and $\Sigma N$ events) and the study of hypernuclei (more than 40
$\Lambda$-hypernuclei, a few $\Lambda\Lambda$-hypernuclei and $\Xi$-hypernuclei), and it is required that hyperonic potentials used should be fitted from the existing $NY$ scattering data (in microscopic approaches) or be consistent with the properties of hypernuclei (in phenomenological approaches).
However, the parameters of the hyperonic interaction cannot be sufficiently well constrained, and there are still many theoretical and experimental ambiguities regarding baryon interactions in the strangeness sector.

The objective of the present study is to confront the phenomenological hyperonic interaction to the recently observed neutron star properties (like mass $M$, radius $R$ and tidal deformability $\Lambda$), and perform Bayesian inference for the hyperon-meson coupling constants from the robust LIGO/Virgo tidal measurement of the GW170817 binary neutron star merger~\citep{2017PhRvL.119p1101A}\footnote{The results of the other possible double-neutron star merger GW190425~\citep{2020ApJ...892L...3A} were found to provide poor EOS constraint~\citep{2021ApJ...913...27L}, mainly due to the intrinsically much smaller tidal deformability and its low signal-to-noise ratio.}
as well as two NICER~mass-radius measurements of pulsars [PSR J0030+0451~\citep{2019ApJ...887L..21R,2019ApJ...887L..24M} and PSR J0740+6620~\citep{2021ApJ...918L..27R,2021ApJ...918L..28M}].  
Previously, such kind of Bayesian analysis on nuclear matter parameters has been performed in e.g., \citet{2020ApJ...905....9T,2022PhRvC.105a5806I,2022arXiv220112552M}, we here focus on hypernuclear matter based on a set of generally-used relativistic mean-field (RMF) Lagrangians.
We consider 18 RMF effective interactions described as both finite-range interactions~\citep{2001PhRvL..86.5647H,2004PhRvC..69c4319L,2005PhRvC..71b4312L,2010PhRvC..81a5803T,2020ChPhC..44g4107W} and zero-range (contact interactions or point couplings PC~\citep{1992PhRvC..46.1757N,2006NuPhA.770....1F,2008PhRvC..78c4318N}, which all treat effectively the in-medium properties of baryon-baryon interaction.
And the properties of nuclear saturation and nuclear symmetry energy are in the empirical ranges given by finite-nuclei and heavy-ion experiments.

The hyperonic interaction parameters within the RMF framework, in terms of an exchange of mesons,
are hyperon-meson coupling constants. Hereafter we define $R_{m Y}=g_{m Y}/g_{m N}$ ($m=\sigma, \omega, \rho, \delta$).
If the SU(6) symmetry is applied for the $\omega$-$\Lambda$-hyperon coupling ($R_{\omega \Lambda}=2/3$), the $\sigma$-$\Lambda$-hyperon coupling can be derived from the extrapolation at $A^{-2/3}=0$ of the experimental binding energy of single-$\Lambda$ hypernuclei: $R_{\sigma \Lambda}\sim0.61$, which corresponds to $U_{\Lambda}(\rho_0)=-30\mev$~\citep{2006PrPNP..57..564H}.
Nevertheless, it is still uncertain to determine the fit to hypernuclear data.
The broken SU(3) flavor symmetry seems inevitable to accommodate the attractive/repulsive nature of different hyperon potentials~\citep[e.g.][]{2017PhRvC..95f5803F} simultaneously and to meet the requirement of the mass measurements of heavy pulsars~\citep[e.g.][]{2012PhRvC..85f5802W}. 
For the present analysis, we then relax the symmetry argument and regard $R_{\sigma \Lambda}$ and $R_{\omega \Lambda}$ as free parameters between 0 and 1, while keeping consistency with the hypernuclear data by imposing an empirical $R_{\sigma \Lambda}$ versus $R_{\omega \Lambda}$ relation from the experimental binding energy.
By doing so, we have reasonably assumed that the hyperon coupling is positive and smaller than the nucleon ones~\citep[e.g.][]{1991PhRvL..67.2414G}.
Since there is not yet sufficient experimental information for the $\Sigma$ and $\Xi$ hypernuclei, we fix the vector-$\Sigma,\Xi$ hyperon couplings based on the above-mentioned SU(6) symmetry, and adopt the scalar-coupling values for $\Sigma,\Xi$ hyperons ($R_{\sigma \Sigma}=0.443$, $R_{\sigma \Xi}$=0.302)~\citep{2014PhLB..734..383V,2013PhRvC..87e5806C} reproducing reasonable single-particle mean-field potentials in symmetric nuclear matter at the saturation density $\rho_0$: $U_{\Sigma}(\rho_0)=+34\mev$~\citep{2017PhRvC..95f5803F} and $U_{\Xi}(\rho_0)=-14\mev$~\citep{2000PhRvC..61e4603K,2016RvMP...88c5004G}.
{As for the isospin-vector scalar channel, following~\citet{2016PhRvC..93b5806D}, we simply take $R_{\delta \Sigma}=R_{\sigma \Sigma}$, $R_{\delta \Xi}=R_{\sigma \Xi}$~\citep{1974PhRvD...9.1613M}.
In the present study, we shall explore the constraints on the $\Lambda$ hyperon coupling constants by the current measurements of the mass, radius, and tidal deformability of neutron stars, and discuss the parameter spaces of hyperon star properties.
We are also interested in providing useful relations between hyperon-star observed properties, supplementing such kinds of EOS-insensitive relations commonly obtained based on neutron star EOSs.

The paper is organized as follows: 
Section 2 briefly introduces the employed RMF model and the descriptions of (hyper)nuclear matter and dense stellar matter, including the discussions on nuclear symmetry energy parameters, the EOS stiffness, etc.
The used neutron star observations and the $R_{\sigma \Lambda}$-$R_{\omega \Lambda}$ relation from hypernuclei data are detailed in Section 3, as well as the Bayesian analysis method.
We present our results and a discussion in Section~4 and summarize our paper in Section~5.

\section{Relativistic EOSs for hypernuclear matter} \label{sec:eos}

\subsection{The RMF model}
 
The RMF model is constructed based on the framework of quantum hadrodynamics~\citep{1974AnPhy..83..491W,1992RPPh...55.1855S}. It has been successfully employed to describe the nuclear structure and is one of the widely-used phenomenological models for studying nuclear many-body systems.
One of the big advantages of the treatment in terms of a relativistic field theory is that it is automatically causal.
Also, we could bypass the problem of unknown hyperon-relevant three-body interaction, which is supposed to induce additional EOS uncertainty in the microscopic calculations based on Schrodinger’s equation.   

 \begin{figure*}
\includegraphics[width = 3.55in]{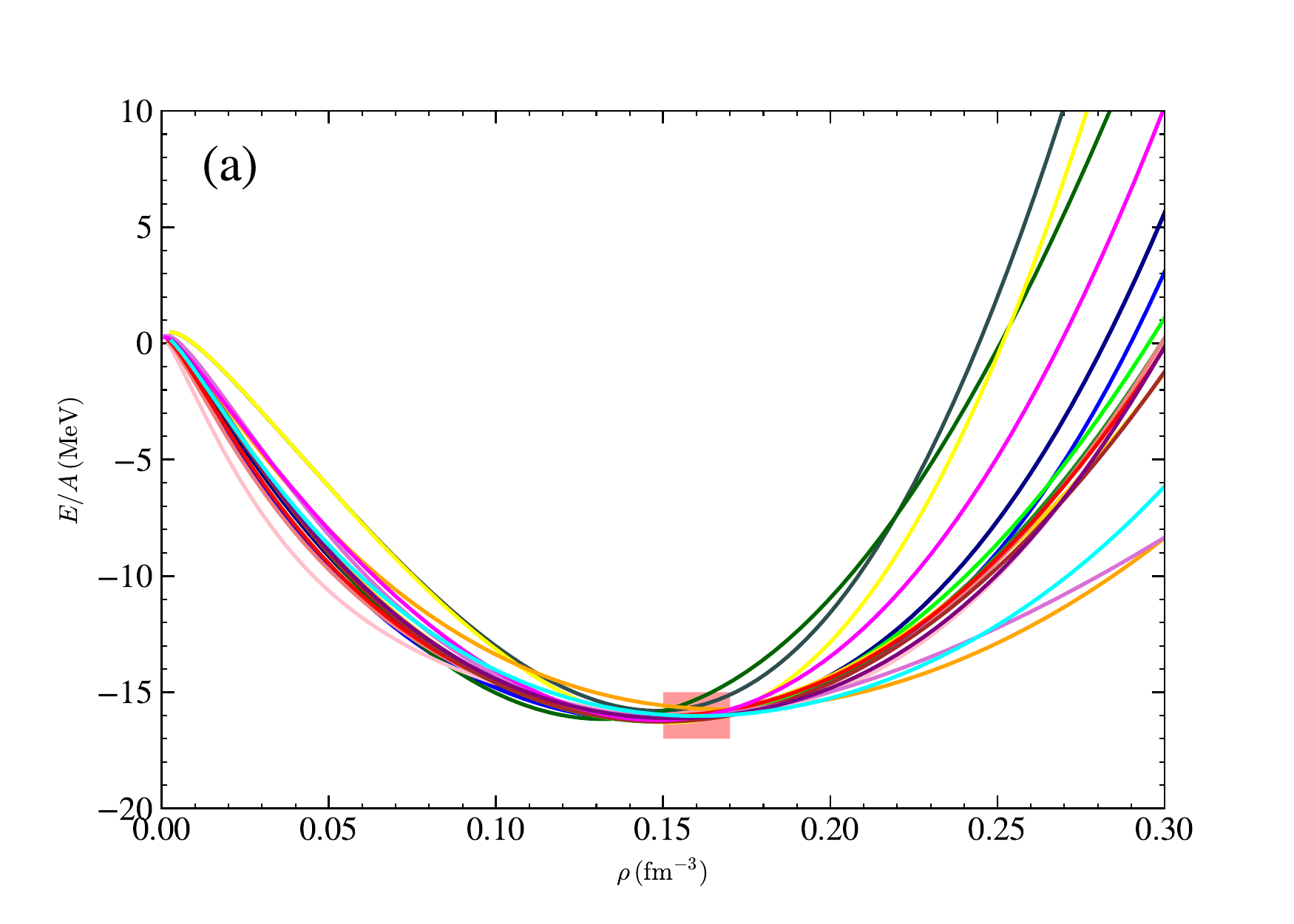}
\includegraphics[width = 3.55in]{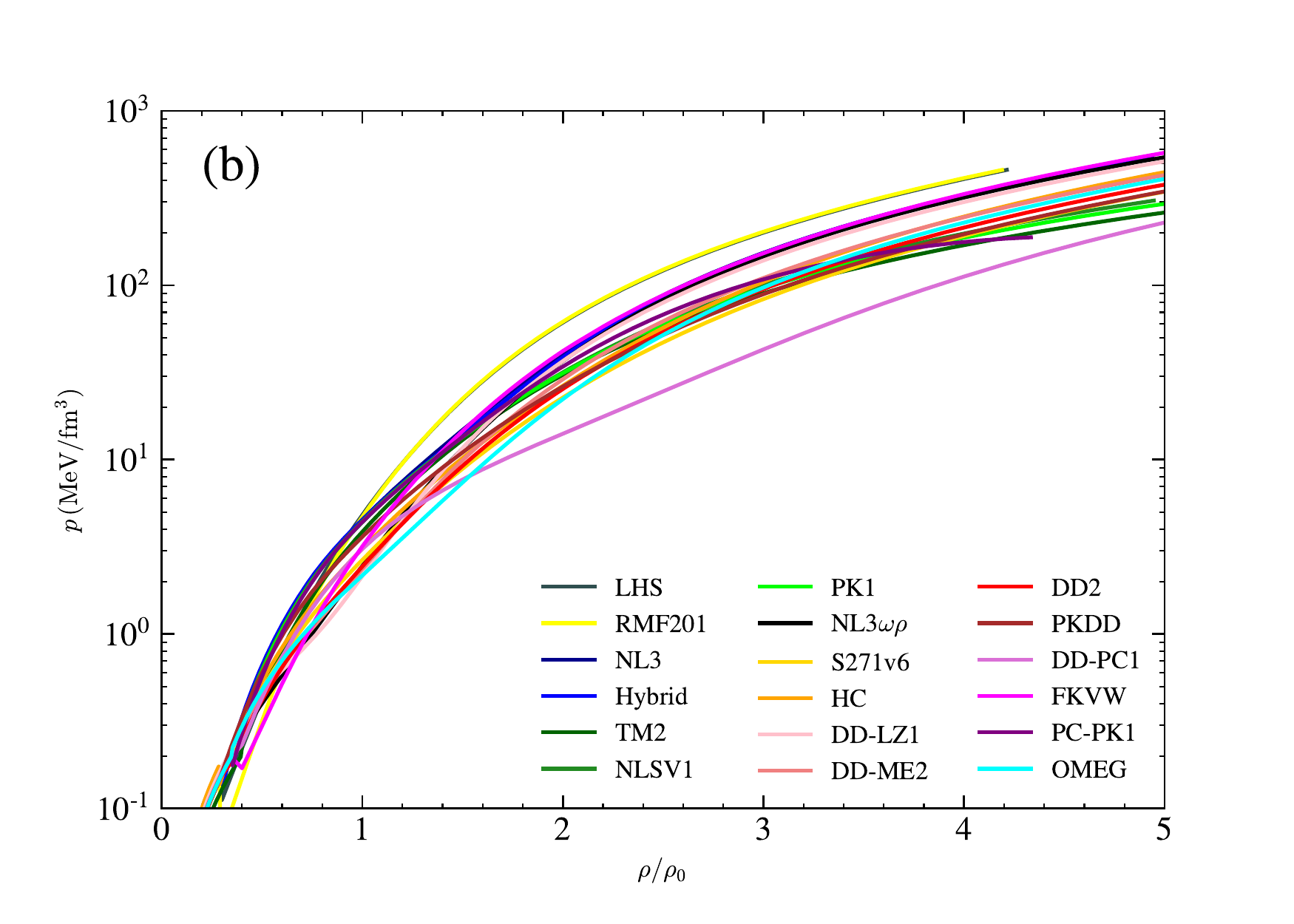}
\includegraphics[width = 3.56in]{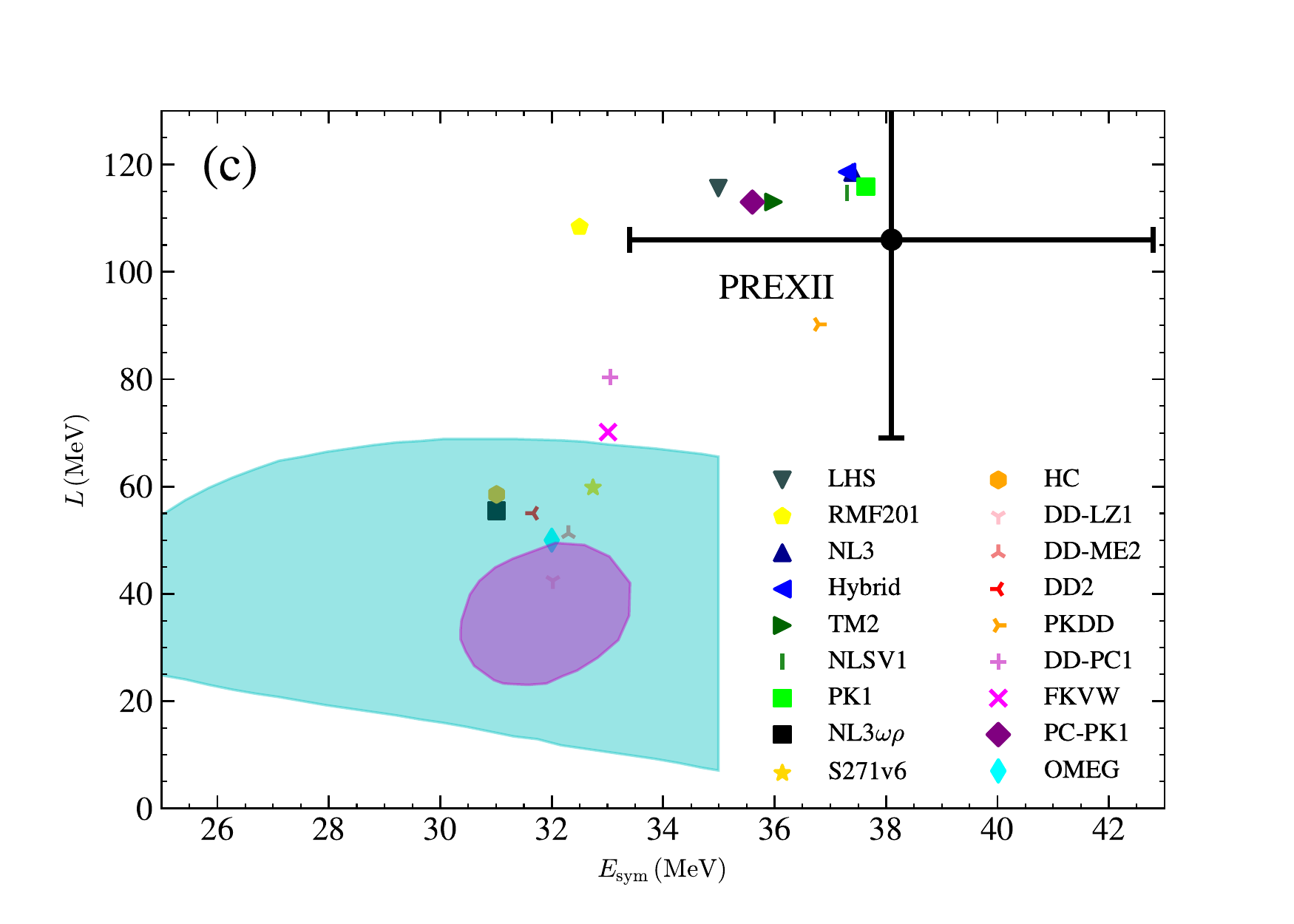}
\includegraphics[width = 3.56in]{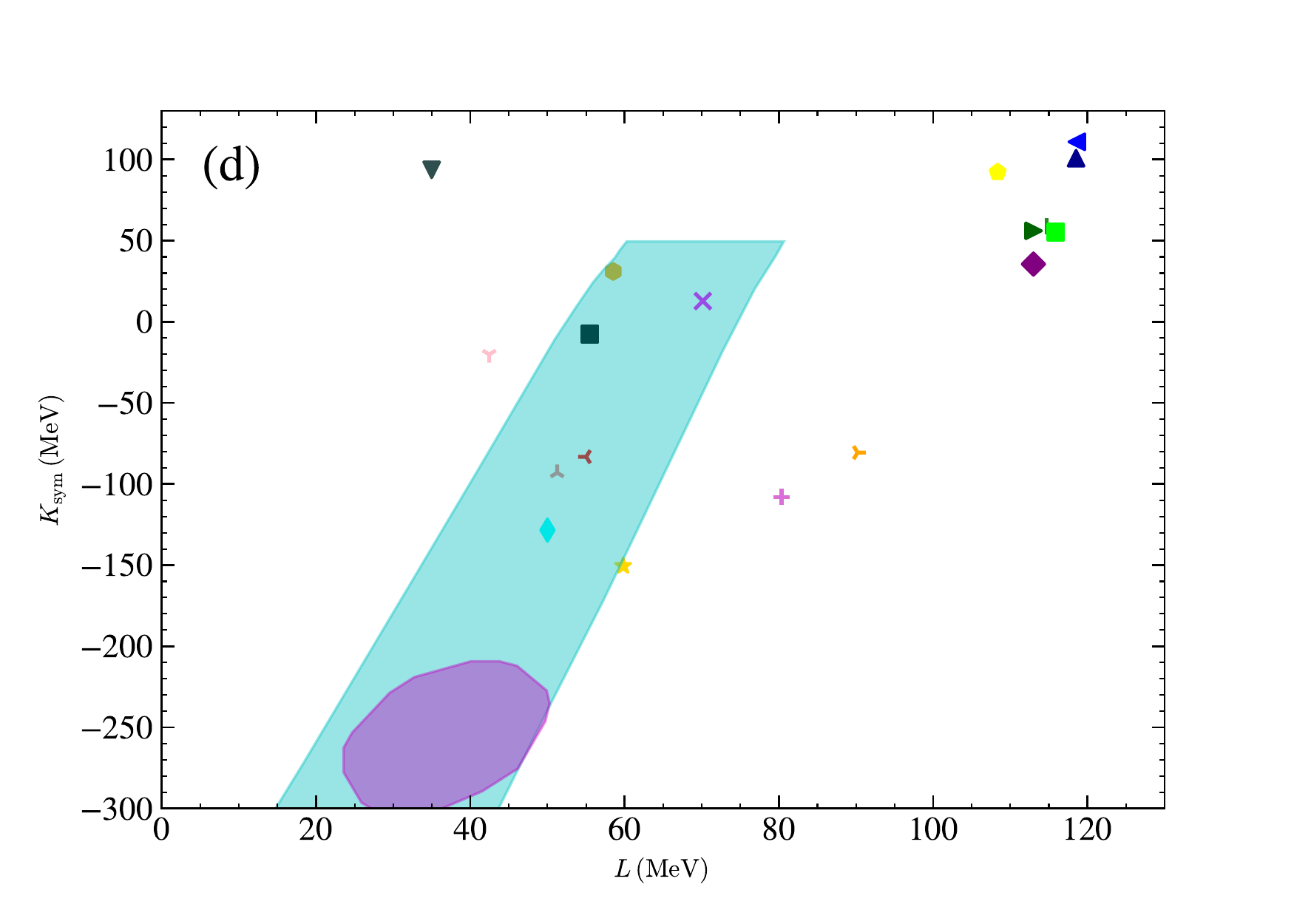} 	\vspace{-0.3cm}
\caption{(Panel a) Energy per nucleon in symmetric nuclear matter as a function of the nucleon density for different models; The pink box represents the empirical regions of $\rho_0\sim0.16\pm0.01\fm3$ and $E/A\sim-16\pm1\mev$;
(Panel b) EOSs for beta-stable matter; 
(Panel c) Values of the symmetry energy $E_{\rm sym}$ and the slope parameter $L$, with the black dots with error bars indicating the recent PREX-II results at $68\%$ credible level~\citep{2021PhRvL.126q2503R}; 
(Panel d) Values of the symmetry energy slope $L$ and the curvature parameter $K_{\rm sym}$.
Also shown in the lower panels are the $95\%$ credible regions from a latest analysis~\citep{2021PhRvC.103f4323N} on neutron skin data of calcium, lead, and tin isotopes, both for uninformative priors in $E_{\rm sym}$, $L$, $K_{\rm sym}$, and for priors incorporating knowledge of the PNM EOS from chiral effective field theory calculations. 
Note that the two original linear Walecka models (LHS and RMF201) are already ruled out because of too large isospin symmetric nuclear matter incompressibility, and the five stiffest EOSs (LHS, RMF201, NL3, Hybrid, TM2) are disfavored by the astrophysical radius constraints (see below in Fig.~\ref{fig:M-R}). 
} 	
\label{fig:EOS}
\end{figure*}

As mentioned in the introduction, we consider various types of RMF effective interactions:
LHS~\citep{1989RPPh...52..439R}, RMF201~\citep{2010PhRvC..82b5203D}, NL3~\citep{1997PhRvC..55..540L}, Hybrid~\citep{2009PhRvC..79e4311P}, TM2~\citep{1994NuPhA.579..557S}, NLSV1~\citep{2000PhRvC..61e4306S}}, PK1~\citep{2004PhRvC..69c4319L}, NL3$\omega\rho$~\citep{2001PhRvL..86.5647H}, S271v6~\citep{2002PhRvC..66e5803H}, HC~\citep{2003PhRvC..68e4318B}, DD-LZ1~\citep{2020ChPhC..44g4107W}, DD-ME2~\citep{2005PhRvC..71b4312L}, DD2~\citep{2010PhRvC..81a5803T}, PKDD~\citep{2004PhRvC..69c4319L}, OMEG~\citep{2022ApJ...929...82M}, 
and the zero-range point-coupling ones without exchanging mesons: DD-PC1~\citep{2008PhRvC..78c4318N}, FKVW~\citep{2006NuPhA.770....1F}, PC-PK1~\citep{2010PhRvC..82e4319Z}.
They also cover all variants of the RMF models reviewed in \citet{2014PhRvC..90e5203D}.
In these models, the in-medium effects of nuclear force are taken into account with effective Lagrangians, in which the relevant parameters, accounting for higher-order many-body effects, are introduced by either including non-linear meson self-interaction terms or assuming an explicit density dependence for the meson-nucleon couplings.
For example, DD-LZ1 is developed with peculiar density-dependent behavior of meson-nucleon couplings guided by the restoration of pseudospin symmetry around the Fermi levels in finite nuclei~\citep{2020ChPhC..44g4107W}.

In meson exchange perspective, the Lagrangian density generally reads as ($B=N, Y$):
$\mathcal{L}$ = $\mathcal{L}_{\rm free}^B + \mathcal{L}^B_{\rm int} + \mathcal{L}_m + \mathcal{L}_{\rm NL}$. 
The terms $\mathcal{L}_{\rm free}^B$, $\mathcal{L}_{\rm int}^B$, $\mathcal{L}_m$ describe, respectively, the free baryons, the baryons interacting with the mesons and the free mesons, while the last one $\mathcal{L}_{\rm NL}$ presents only in the nonlinear version to involve the self-interaction and nonlinear mixing of mesons. 
In the zero-range PC models, instead of meson exchange diagram, the baryon effective interaction in medium is treated by several local four-point (contact) terms between baryons~\citep[see discussions in e.g.,][]{2010PhRvC..82e4319Z,2019AIPC.2127b0020S}. Correspondingly, their Lagrangian density could be written as: $\mathcal{L}_{\rm PC} = \mathcal{L}_{\rm free}^B + \mathcal{L}_{\rm 4f}^B + \mathcal{L}_{\rm hot}^B$, where only baryons' degrees of freedom or their densities involve. In general, the terms in these RMF models are given by:
\begin{equation}
\begin{aligned}
& \mathcal{L}^B_{\rm free} =\sum\limits_{B}\bar{\psi}_B \left[i\gamma_{\mu}\partial^{\mu}-m_B \right] \psi_B \ ;
\end{aligned}
\end{equation}
\begin{equation}
\begin{aligned}
& \mathcal{L}^B_{\rm int} =\sum\limits_{B}\bar{\psi}_B \left[-g_{\sigma B}\sigma 
- g_{\omega B}\gamma_\mu\omega^\mu - g_{\rho B} \gamma_\mu\vec{\rho}^\mu\cdot\vec{\tau} \right] \psi_B \ ;
\end{aligned}
\end{equation}
\begin{equation}
\begin{aligned}
\mathcal{L}_m=&
-\frac{1}{2} m_\sigma^2\sigma^2
+\frac{1}{2} m_\omega^2 \omega_\mu\omega^\mu
+\frac{1}{2} m_\rho^2\vec{\rho}_\mu\cdot\vec{\rho}^\mu  \\ 
&+\frac{1}{2}\partial_\mu\sigma\partial^\mu\sigma
-\frac{1}{4}\Omega_{\mu\nu}\Omega^{\mu\nu}
-\frac{1}{4}\vec{ R}_{\mu\nu}\cdot\vec{ R}^{\mu\nu} \ ;
\end{aligned}
\end{equation}
\begin{equation}
\begin{aligned}
\mathcal{L}_{\rm NL} =&-\frac{1}{3}g_2 \sigma^3 - \frac{1}{4}g_3 \sigma^4 + \frac{1}{4}c_3(\omega_\mu\omega^\mu)^2   \\ 
&+\Lambda_v(g_{\omega B}^2\omega_{\mu}\omega^{\mu})(g_{\rho B}^2\rho_{\mu}\rho^{\mu}) \ ;
\end{aligned}
\end{equation}
\begin{equation}
\begin{aligned}
\mathcal{L}_{\rm 4f}^B=&-\frac{1}{2}\sum_B \alpha_S^{NB}(\rho)(\bar{\psi}_N\psi_N)(\bar{\psi}_B\psi_B)
\\ 
&-\frac{1}{2}\sum_B \alpha_V^{NB}(\rho)(\bar{\psi}_N\gamma_{\nu}\psi_N)(\bar{\psi}_B\gamma_{\nu}\psi_B)
\\ 
&-\frac{1}{2}\sum_B \alpha_{TS}^{NB}(\rho)(\bar{\psi}_N\vec{\tau}\psi_N)(\bar{\psi}_B\vec{\tau}\psi_B)\\
&-\frac{1}{2}\sum_B \alpha_{TV}^{NB}(\rho)(\bar{\psi}_N\vec{\tau}\gamma_{\nu}\psi_N)(\bar{\psi}_B\vec{\tau}\gamma_{\nu}\psi_B)\ ;
\end{aligned}
\end{equation}
\begin{equation}
\begin{aligned}
\mathcal{L}_{\rm hot}^B=&-\frac{1}{3}\beta_S^{NN}(\bar{\psi}_N\psi_N)^3-\frac{1}{4}\gamma_S^{NN}(\bar{\psi}_N\psi_N)^4 \\
&-\frac{1}{4}\gamma_V^{NN}[(\bar{\psi}_N\gamma_\mu\psi_N)][(\bar{\psi}_N\gamma^\mu\psi_N)]^2\ ,
\end{aligned}
\end{equation}
where $\psi_B$ describes baryon $B$, $\sigma$ is scalar-isoscalar meson field, $\omega^{\mu}$ and $\vec{\rho}^\mu$ stand for the vector-isoscalar and isovector fields, respectively. The vector meson field tensors $\Omega^{\mu\nu}$ and $\vec{ R}_{\mu\nu}$ are defined by $V_{\mu\nu}= \partial_{\mu} V_{\nu}-\partial_{\nu} V_{\mu}$. 
For the PC models, the coupling strengths $\alpha_S$, $\alpha_V$, $\alpha_{TS}$ and $\alpha_{TV}$ introduce various tensor structure of interactions in spin-isospin space.
$\mathcal{L}_{\rm 4f}^B$ is the four-fermion interactions. The higher-order terms involving more than four fermions are introduced in $\mathcal{L}_{\rm hot}^B$, reflecting the effects of medium dependence.
At the mean-field level, the many-body state is built up as a Slater determinant from single-particle wave functions, i.e., four-component Dirac spinors.
The Klein-Gordon equations for the meson fields and the Dirac equations for the baryon field are solved self-consistently in the RMF approximation, where the meson-field operators are replaced by their expectation values in the ground state. 
For the PC models, the procedure is simplified further as only the baryon field exists.
The nucleon couplings in our set of EOS models are determined to reproduce the binding energies, charge radii, neutron radii of selected nuclei, as well as the properties of symmetric and asymmetric nuclear matter.
In particular, NL3$\omega\rho$ includes a mixed isoscalar-isovector term, which is varied to change the density dependence of the symmetry energy~\citep{2001PhRvL..86.5647H}, and we take it as a representative model of one of the stiffest EOSs in the literature. 

In particular, the saturation single-$\Lambda$ potential in symmetric nuclear matter is determined from the scalar and vector potentials, $U_{\Lambda} = g_{\sigma\Lambda}\sigma+g_{\omega\Lambda}\omega_0$ for meson-exchange models and $U_{\Lambda} =\alpha_S^{(N\Lambda)}\rho_S+\alpha_V^{(N\Lambda)}\rho_V$ for PC models~\citep{2012PhRvC..85a4306T}, with $\rho_S$ being the scalar density and $\rho_V$ the time-like component of the vector density (namely, $\rho_V=\rho$).
See e.g.~\citet{2014PhRvC..90e5203D} for a general review on the application of the RMF model to properties of nuclear matter. 
There appears a linear behaviors between the ratios $R_{\sigma\Lambda}$ and $R_{\omega\Lambda}$ in both nonlinear~\citep{2000PhRvC..61f4309K} and density-dependent~\citep{2021PhRvC.104e4321R} classes of the RMF model. 
Such linear relation has been recently fitted~\citep{2021PhRvC.104e4321R} from the known $\Lambda$ separation energy of hypernuclei, with the analysis of the fitting errors provided. 
We shall make use of this relation in our analysis and denote as the likelihood function $P_{\rm NUCL}$ (see below in Sec.~\ref{sec:analysis}).  

\subsection{Relativistic EOSs for nuclear and hypernuclear matter}
\label{sec:eos2}

By construction, the RMF effective interactions are well behaved close to the nuclear saturation density $\rho_0$ and moderate values of the isospin asymmetry. See Fig.~\ref{fig:EOS}(a) for the energy per particle for symmetric nuclear matter plotted as a function of baryon density for all 18 chosen effective interactions.
They give an acceptable saturation point, which means a saturation density in the
interval of $\rho_0\sim0.15$-$0.17\fm3$ and the corresponding energy per particle at saturation in the interval of $E/A\sim-(17$-$15)\mev$;
And the calculated incompressibility at saturation is compatible with the constraints of $190$-$270$ MeV~\citep{2014PhRvC..90e5203D} from the analysis on giant monopole and dipole resonances, except the original linear Walecka models of LHS~\citep{1989RPPh...52..439R} and RMF201~\citep{2010PhRvC..82b5203D}, having extreme values of 577.84, 548.10 MeV, respectively, which we only include them for comparison. 
Nevertheless, the symmetry energy parameters ($E_{\rm sym}$, $L$, $K_{\rm sym}$) spread the two-dimensional plots of $L$ versus $E_{\rm sym}$ and $L$ versus $K_{\rm sym}$ in the lower panels of Fig.~\ref{fig:EOS}.

$E_{\rm sym}$, $L$, $K_{\rm sym}$ are usually refereed to the magnitude, slope, and curvature of the symmetry energy at saturation density: $E_{\rm sym} (\rho)=E_{\rm sym}+\chi L+ \frac{1}{2}\chi^2K_{\rm sym}+...$ with $\chi=(\rho-\rho_0)/(3\rho_0)$. Therefore 18 RMF effective interactions predict different EOSs for asymmetric matter and pure neutron matter (PNM).
Since microscopic calculations of PNM are very accurate, they can serve as a good reference for the phenomenological models.
In the lower panels of Fig.~\ref{fig:EOS} of the symmetry energy parameters, we also include the $95\%$ credible regions from a latest analysis~\citep{2021PhRvC.103f4323N} on neutron skin data (one of the most accessible measurements of neutron-rich environments in the Laboratory), both for uninformative priors in $E_{\rm sym}$, $L$, $K_{\rm sym}$ over their empirical ranges, and for priors incorporating knowledge of the PNM EOS from chiral effective field theory calculations (see the quoted reference and references therein for more details).

\begin{figure}
\includegraphics[width = 0.96\linewidth]{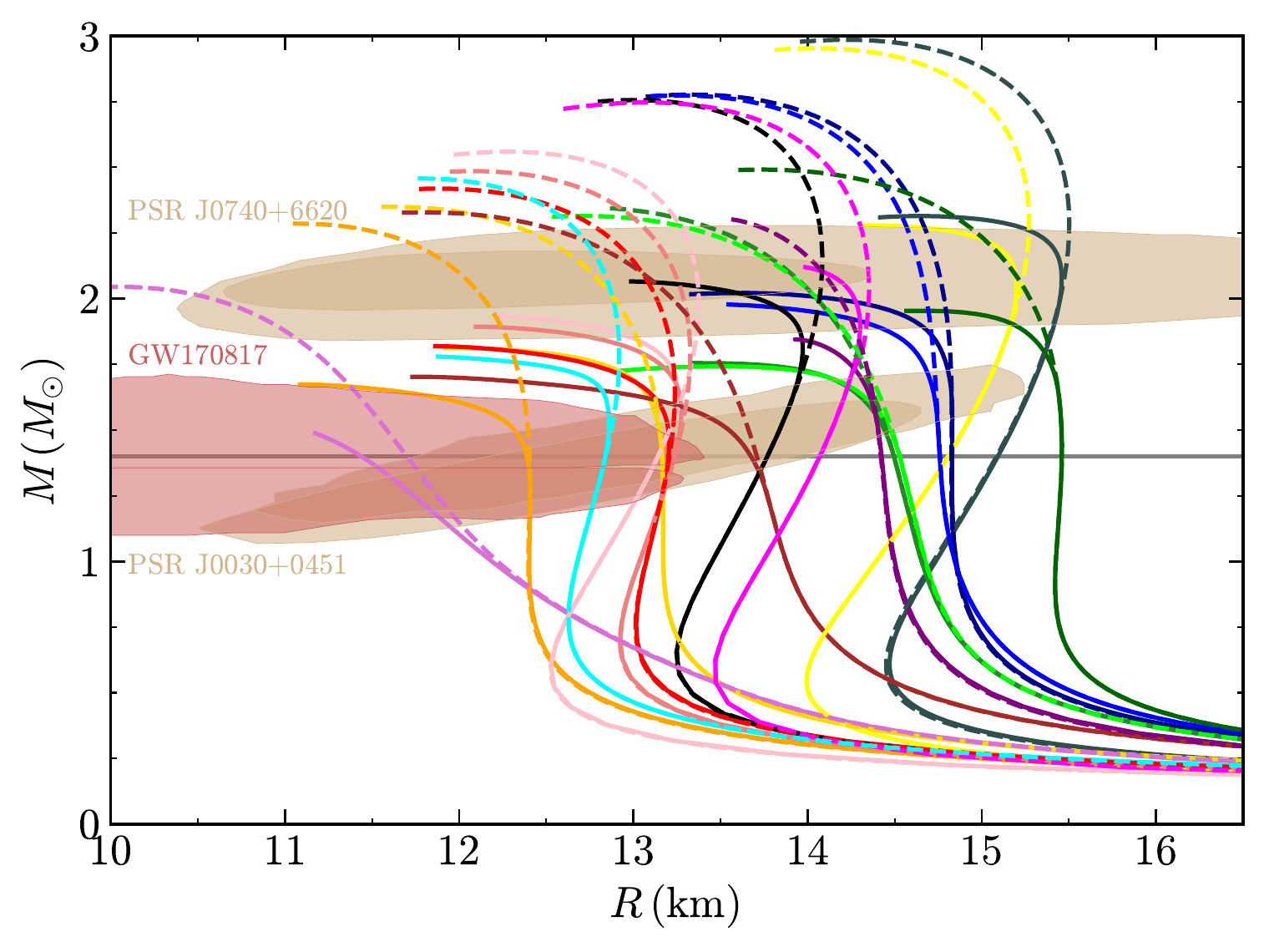} 
\includegraphics[width = 0.96\linewidth]{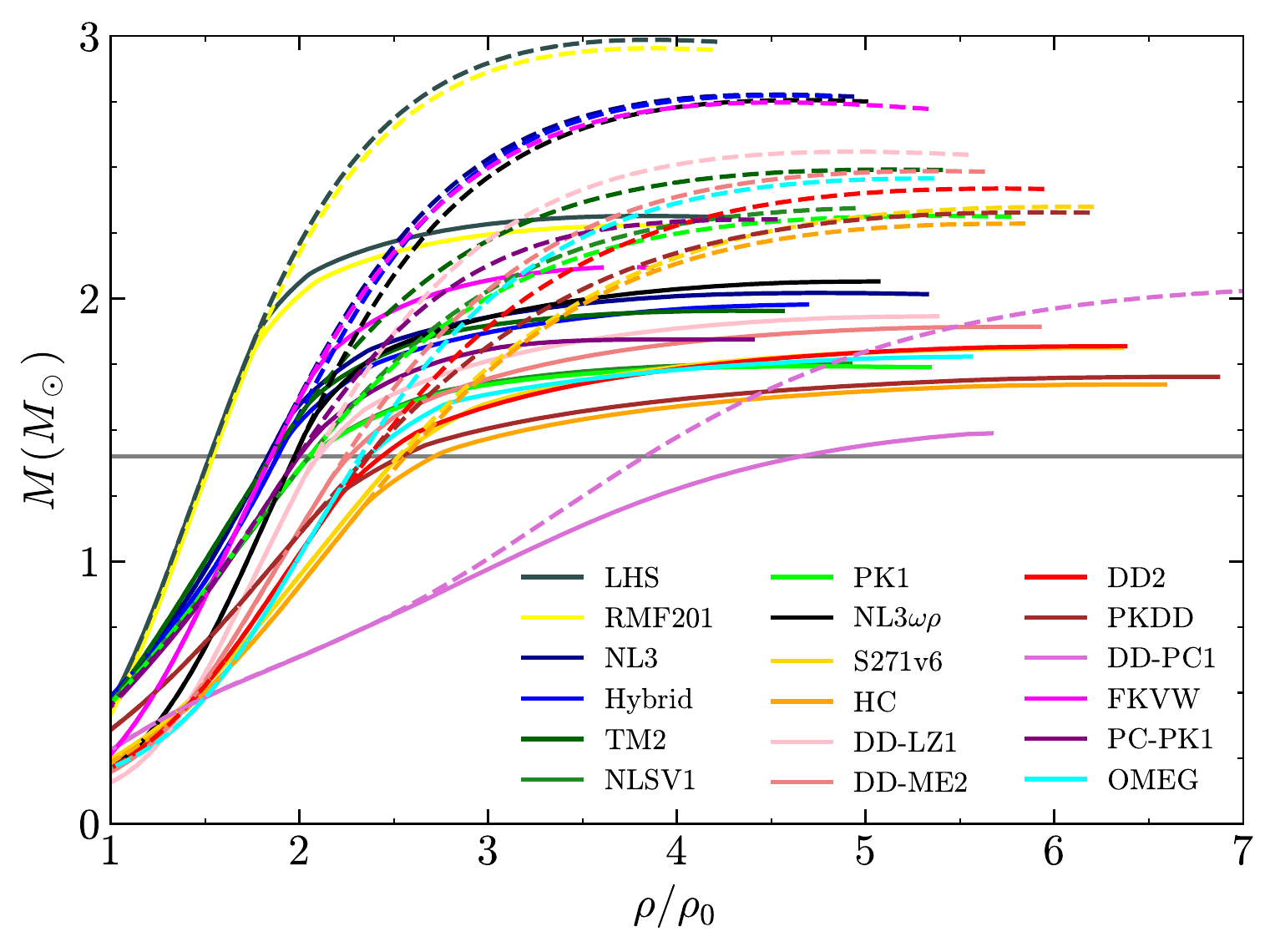} \vspace{-0.3cm}
\caption{Gravitational mass as a function of the stellar radius (upper panel) and the central density (scaled by the nuclear saturation density $\rho_0$; lower panel) in the case of pure neutron stars (dashed curves) and hyperon stars (solid curves) for the considered EOS models.
The mass-radius measurements of the NICER mission for PSR J0030+0451~\citep{2019ApJ...887L..24M,2019ApJ...887L..21R} and PSR J0740+6620~\citep{2021ApJ...918L..27R,2021ApJ...918L..28M} are also shown, along with the binary tidal deformability measurement from GW170817 by LIGO/Virgo~\citep{2017PhRvL.119p1101A}.
The horizontal lines in the two panels indicate $1.4\Msun$. 
For the hyperon star plot, the scalar couplings are fixed as $R_{\sigma\Lambda}=0.611$, $R_{\sigma\Sigma}=0.443$, and $R_{\sigma\Xi}=0.303$. See text for details. 
} \vspace{-0.3cm}
\label{fig:M-R}
\end{figure}

In Fig.~\ref{fig:EOS}(b), we report the pressure versus density relations, i.e., the EOS, of the beta-stable matter for the study of neutron star observations.
Electrons are treated as free ultrarelativistic gas, whereas the muons are relativistic.
The EOS for beta-stable matter can be used in the Tolman-Oppenheimer-Volkoff equations to compute the neutron star mass and radius as a function of the central density.
The mass as functions of the radius and the central density of neutron stars and hyperon stars are displayed in Fig.~\ref{fig:M-R} with dashed and solid curves, respectively.  
As an illustrative example, we show the hyperon star results calculated at $R_{\sigma\Lambda}=0.611$, $R_{\sigma\Sigma}=0.443$, and $R_{\sigma\Xi}=0.303$ for the scalar couplings. 
We also show in the mass-radius plot the mass and radius constraints obtained from the GW170817 tidal deformability measurement by LIGO/Virgo ~\citep{2017PhRvL.119p1101A} and from the NICER measurements for PSR J0030+0451~\citep{2019ApJ...887L..24M,2019ApJ...887L..21R} and PSR J0740+6620~\citep{2021ApJ...918L..27R,2021ApJ...918L..28M}.
To describe the structure of the crust, we employ the quantal calculations of \citet{1973NuPhA.207..298N} for the medium-density regime
($0.001\;\mathrm{fm}^{-3}<\rho<0.08\;\mathrm{fm}^{-3}$),
and follow the formalism developed in \citet{1971ApJ...170..299B} for the outer crust ($\rho<0.001\;\mathrm{fm}^{-3}$).

\begin{table*}
	\centering
	\caption{Most probable intervals of $R_{\sigma\Lambda}$ and $R_{\omega\Lambda}$ ($68\%$ credible intervals) of four different tests (see details in Sec.~\ref{sec:analysis}) for 17 considered relativistic EOSs (see details in Sec.~\ref{sec:eos}).  The softest DD-PC1 is not included because it can not support a $\gtrsim 2.07\Msun$ pure neutron star~\citep{2021ApJ...915L..12F}.}	\label{tab:coupling} 	\vspace{-0.3cm}
	\setlength{\tabcolsep}{0.8pt}
\renewcommand\arraystretch{1.5}
  \begin{ruledtabular}
	\begin{tabular}{cccccccccc} 
	\multicolumn{2}{c}{\multirow{2}{*}{EOS}}
	&\multicolumn{2}{c}{\makebox[0.09\textwidth][c]{+NICER}}
	&\multicolumn{2}{c}{\makebox[0.09\textwidth][c]{+NICER +NUCL}} &\multicolumn{2}{c}{\makebox[0.09\textwidth][c]{+NICER +GW170817}} &\multicolumn{2}{c}{\makebox[0.09\textwidth][c]{+NICER +GW170817 +NUCL}}\\
    \multicolumn{2}{c}{}&$R_{\sigma\Lambda}$&$R_{\omega\Lambda}$&$R_{\sigma\Lambda}$&$R_{\omega\Lambda}$&$R_{\sigma\Lambda}$&$R_{\omega\Lambda}$&$R_{\sigma\Lambda}$&$R_{\omega\Lambda}$\\
    \hline
    \multicolumn{2}{c}{LHS}   &$0.821^{+0.125}_{-0.463}$ &$0.755^{+0.073}_{-0.155}$ &$0.865^{+0.074}_{-0.208}$ &$0.658^{+0.130}_{-0.194}$ &$0.941^{+0.035}_{-0.048}$ &$0.763^{+0.034}_{-0.028}$ &$0.658^{+0.163}_{-0.251}$ &$0.752^{+0.049}_{-0.095}$\\\hline
    \multicolumn{2}{c}{RMF201}&$0.760^{+0.186}_{-0.520}$ &$0.759^{+0.081}_{-0.224}$ &$0.658^{+0.172}_{-0.249}$ &$0.672^{+0.138}_{-0.215}$ &$0.949^{+0.032}_{-0.056}$ &$0.769^{+0.035}_{-0.028}$ &$0.842^{+0.090}_{-0.250}$ &$0.754^{+0.061}_{-0.136}$\\\hline    
    \multicolumn{2}{c}{NL3}   &$0.424^{+0.330}_{-0.293}$ &$0.746^{+0.156}_{-0.261}$ &$0.681^{+0.171}_{-0.247}$ &$0.768^{+0.136}_{-0.214}$ &$0.399^{+0.379}_{-0.291}$ &$0.794^{+0.128}_{-0.216}$ &$0.765^{+0.130}_{-0.191}$ &$0.840^{+0.101}_{-0.163}$\\\hline
    \multicolumn{2}{c}{Hybrid}&$0.363^{+0.381}_{-0.265}$ &$0.807^{+0.132}_{-0.276}$ &$0.750^{+0.130}_{-0.179}$ &$0.865^{+0.096}_{-0.157}$ &$0.305^{+0.388}_{-0.217}$ &$0.764^{+0.143}_{-0.254}$ &$0.777^{+0.118}_{-0.181}$ &$0.869^{+0.090}_{-0.147}$\\\hline
    \multicolumn{2}{c}{TM2}   &$0.311^{+0.330}_{-0.221}$ &$0.751^{+0.179}_{-0.494}$ &$0.736^{+0.145}_{-0.201}$ &$0.856^{+0.102}_{-0.193}$ &$0.323^{+0.487}_{-0.237}$ &$0.784^{+0.158}_{-0.300}$ &$0.772^{+0.137}_{-0.239}$ &$0.870^{+0.086}_{-0.204}$\\\hline
    \multicolumn{2}{c}{NLSV1} &$0.252^{+0.285}_{-0.183}$ &$0.756^{+0.167}_{-0.281}$ &$0.688^{+0.117}_{-0.227}$ &$0.863^{+0.100}_{-0.199}$ &$0.247^{+0.279}_{-0.177}$ &$0.744^{+0.182}_{-0.259}$ &$0.689^{+0.122}_{-0.225}$ &$0.866^{+0.100}_{-0.206}$\\\hline
    \multicolumn{2}{c}{PK1}   &$0.254^{+0.273}_{-0.185}$ &$0.756^{+0.172}_{-0.250}$ &$0.687^{+0.139}_{-0.222}$ &$0.869^{+0.099}_{-0.216}$ &$0.248^{+0.271}_{-0.170}$ &$0.754^{+0.176}_{-0.247}$ &$0.683^{+0.130}_{-0.220}$ &$0.867^{+0.101}_{-0.222}$\\\hline
    \multicolumn{2}{c}{NL3$\omega\rho$}&$0.384^{+0.393}_{-0.280}$ &$0.773^{+0.147}_{-0.247}$ &$0.690^{+0.163}_{-0.208}$ &$0.759^{+0.131}_{-0.176}$ &$0.420^{+0.448}_{-0.294}$ &$0.777^{+0.127}_{-0.269}$ &$0.712^{+0.157}_{-0.215}$ &$0.778^{+0.121}_{-0.183}$\\\hline
    \multicolumn{2}{c}{S271v6}&$0.287^{+0.290}_{-0.207}$ &$0.775^{+0.158}_{-0.232}$ &$0.750^{+0.105}_{-0.144}$ &$0.886^{+0.080}_{-0.128}$ &$0.304^{+0.286}_{-0.083}$ &$0.782^{+0.157}_{-0.230}$ &$0.740^{+0.118}_{-0.161}$ &$0.884^{+0.083}_{-0.147}$\\\hline
    \multicolumn{2}{c}{HC}    &$0.266^{+0.253}_{-0.192}$ &$0.517^{+0.316}_{-0.370}$ &$0.733^{+0.110}_{-0.156}$ &$0.902^{+0.070}_{-0.134}$ &$0.266^{+0.304}_{-0.189}$ &$0.783^{+0.157}_{-0.226}$ &$0.737^{+0.106}_{-0.160}$ &$0.902^{+0.072}_{-0.134}$\\\hline
    \multicolumn{2}{c}{DD-LZ1}&$0.298^{+0.321}_{-0.218}$ &$0.775^{+0.152}_{-0.251}$ &$0.769^{+0.122}_{-0.190}$ &$0.871^{+0.083}_{-0.148}$ &$0.327^{+0.381}_{-0.223}$ &$0.792^{+0.142}_{-0.254}$ &$0.772^{+0.128}_{-0.177}$ &$0.870^{+0.087}_{-0.139}$\\\hline
    \multicolumn{2}{c}{DD-ME2}&$0.275^{+0.337}_{-0.192}$ &$0.771^{+0.167}_{-0.299}$ &$0.770^{+0.120}_{-0.172}$ &$0.885^{+0.078}_{-0.137}$ &$0.267^{+0.345}_{-0.188}$ &$0.776^{+0.160}_{-0.237}$ &$0.767^{+0.128}_{-0.168}$ &$0.883^{+0.079}_{-0.124}$\\\hline
    \multicolumn{2}{c}{DD2}   &$0.292^{+0.346}_{-0.205}$ &$0.775^{+0.163}_{-0.252}$ &$0.783^{+0.121}_{-0.173}$ &$0.901^{+0.071}_{-0.135}$ &$0.305^{+0.392}_{-0.221}$ &$0.785^{+0.153}_{-0.276}$ &$0.789^{+0.119}_{-0.157}$ &$0.900^{+0.069}_{-0.120}$\\\hline
    \multicolumn{2}{c}{PKDD}  &$0.267^{+0.347}_{-0.185}$ &$0.806^{+0.140}_{-0.244}$ &$0.820^{+0.095}_{-0.153}$ &$0.930^{+0.051}_{-0.090}$ &$0.282^{+0.420}_{-0.210}$ &$0.813^{+0.136}_{-0.248}$ &$0.835^{+0.102}_{-0.147}$ &$0.932^{+0.047}_{-0.083}$\\\hline
    \multicolumn{2}{c}{FKVW}  &$0.327^{+0.343}_{-0.236}$ &$0.677^{+0.217}_{-0.260}$ &$0.647^{+0.196}_{-0.250}$ &$0.706^{+0.171}_{-0.211}$ &$0.353^{+0.356}_{-0.240}$ &$0.696^{+0.272}_{-0.203}$ &$0.658^{+0.177}_{-0.254}$ &$0.716^{+0.158}_{-0.217}$\\\hline
    \multicolumn{2}{c}{PC-PK1}&$0.283^{+0.310}_{-0.210}$ &$0.701^{+0.215}_{-0.134}$ &$0.650^{+0.150}_{-0.205}$ &$0.770^{+0.147}_{-0.214}$ &$0.282^{+0.319}_{-0.211}$ &$0.703^{+0.212}_{-0.139}$ &$0.651^{+0.148}_{-0.208}$ &$0.771^{+0.146}_{-0.215}$\\\hline
    \multicolumn{2}{c}{OMEG}&$0.272^{+0.298}_{-0.194}$ &$0.778^{+0.156}_{-0.244}$ &$0.726^{+0.117}_{-0.171}$ &$0.880^{+0.089}_{-0.153}$ &$0.273^{+0.275}_{-0.188}$ &$0.775^{+0.163}_{-0.242}$ &$0.731^{+0.119}_{-0.167}$ &$0.889^{+0.082}_{-0.152}$\\
	\end{tabular}
\end{ruledtabular}
\end{table*}

\begin{figure*}
\includegraphics[width = 0.49\linewidth]{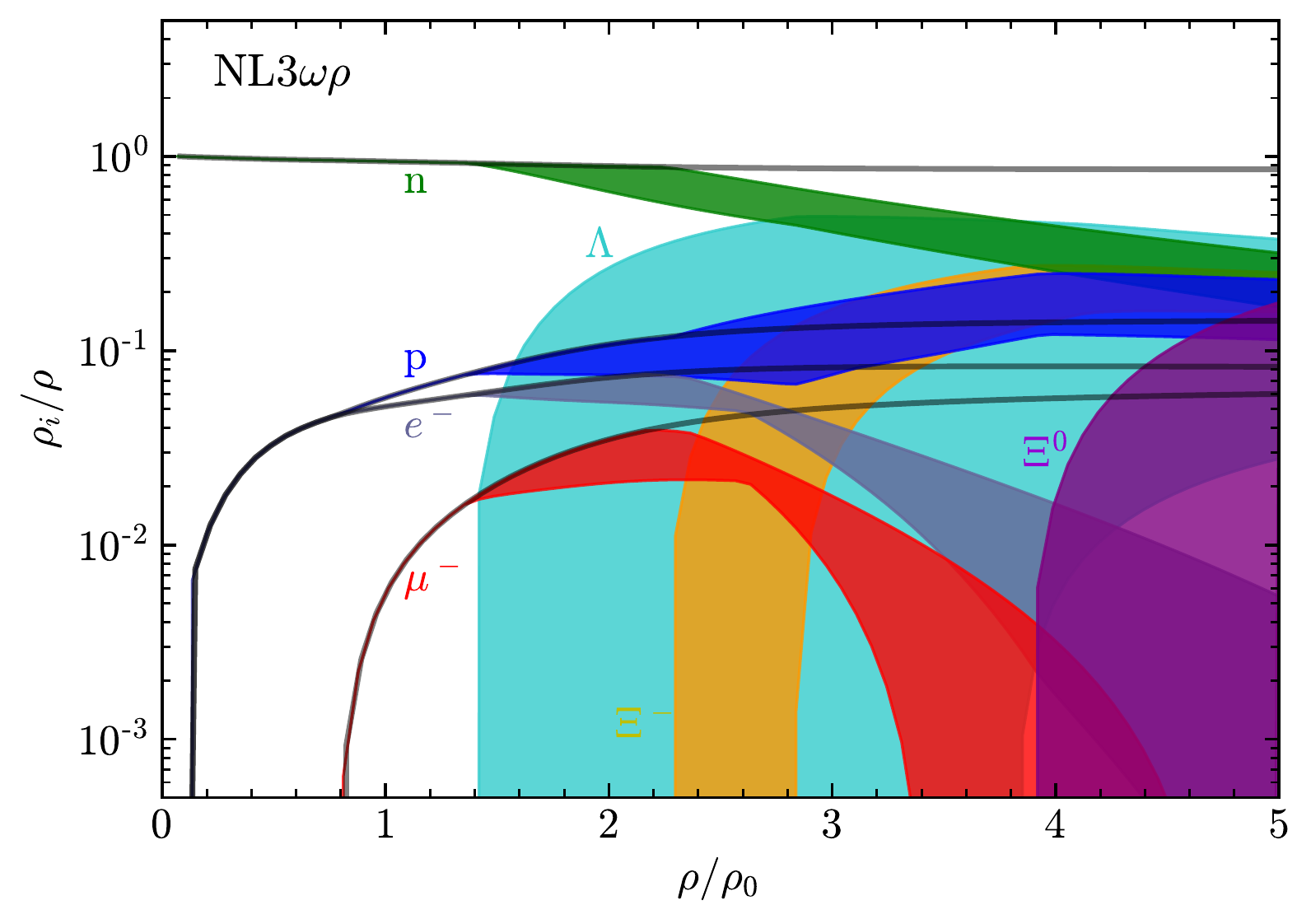}
\includegraphics[width = 0.49\linewidth]{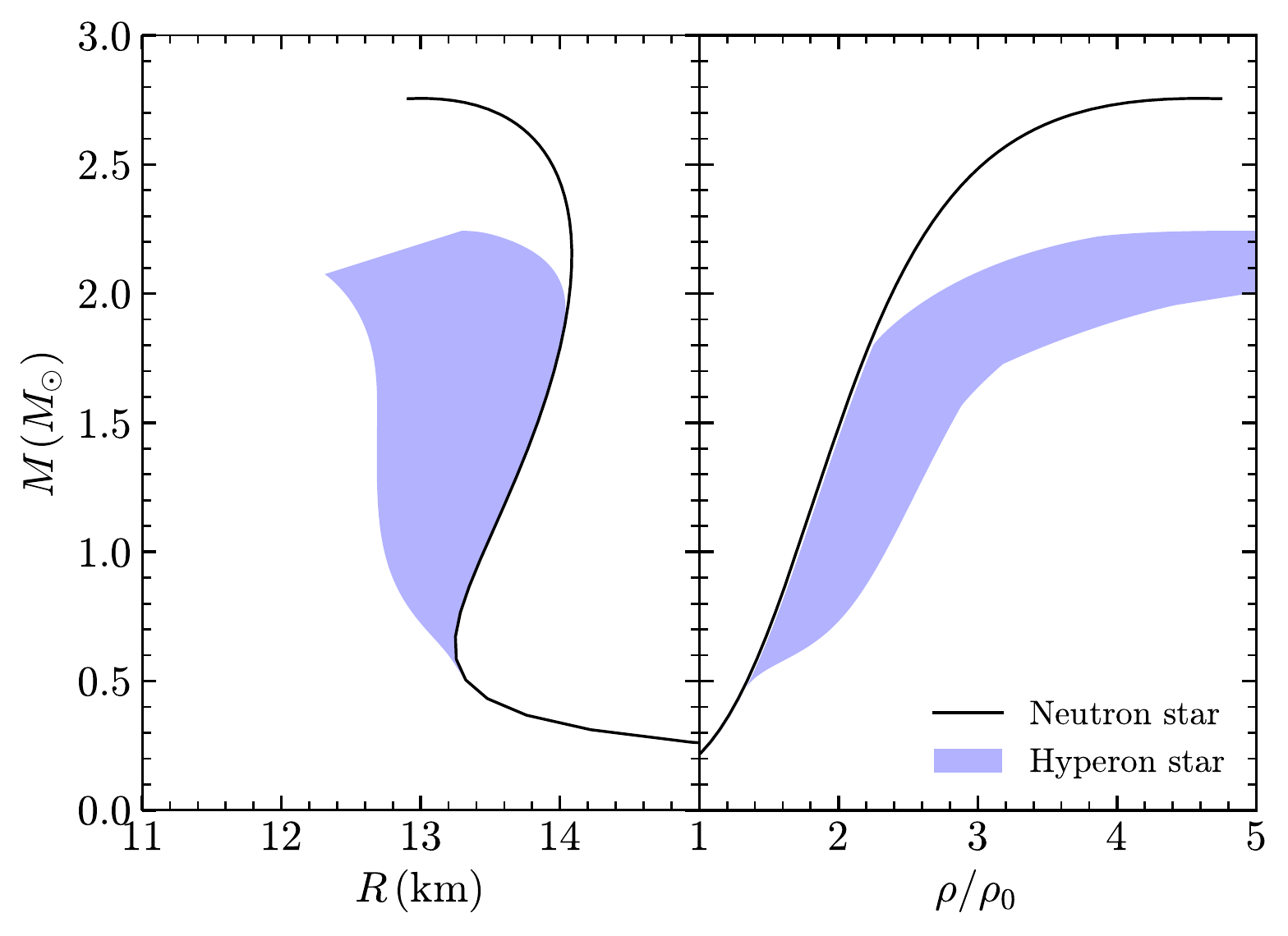} \vspace{-0.3cm}
\caption{(Left panel) Particle fractions of beta-stable hypernuclear matter as a function of the baryon density (scaled by the nuclear saturation density $\rho_0$), with the most probable hyperonic interaction from the joint +NICER+GW170817+NUCL analysis (at $68\%$ credible level) for the representative NL3$\omega\rho$ model; (Right panel) The corresponding most probable mass versus radius \& density relations of hyperon stars, compared to those of neutrons stars. 
} 
\label{fig:hyperonstar}
\end{figure*}

In Fig.~\ref{fig:M-R}, we see that the softest DD-PC1 can not support a $\gtrsim 2.07\Msun$ pure neutron star~\citep{2021ApJ...915L..12F}, while the five stiffest EOSs (LHS, RMF201, NL3, Hybrid, TM2) are not supported by both radius constraints from LIGO/Vigo and NICER due to either too large symmetric nuclear matter incompressibility (in the cases of LHS and RMF201) or too large slope of the symmetry energy at saturation (in the cases of  NL3, Hybrid, TM2).
For the set of 18 RMF EOS models shown in the figure, it is seen that it is not necessarily a stiff EOS (resulting in a large maximum mass) corresponds to a large stellar radius~\citep{2007PhR...442..109L}:
The neutron star radius is controlled mainly by the density dependence of the nuclear symmetry energy around the nuclear saturation density $\rho_0$, while the maximum mass is a reflection of the EOS stiffness at several times the nuclear saturation density (for example, $\gtrsim5\rho_0$). 
The fact that the EOS models of NLSV1, PK1 and PC-PK1 predict large stellar radii but small maximum masses, with the corresponding hyperon stars felling out of the $90\%$ credible regions of the PSR J0740+6620 mass measurements~\citep{2021ApJ...918L..27R,2021ApJ...918L..28M}, is because their pressure versus density relation are relatively soft at supra-nuclear densities (as shown in Fig.~\ref{fig:EOS} (b)).
Moreover, except NLSV1, PK1 and TM2, accompanied by the decrease of the maximum mass with the addition of the hyperon degrees of freedom, the maximum central density of neutron stars increases. See again later in Table~\ref{tab:hyperonstar} and Fig.~\ref{fig:hyperonstar} for the comparison of the central densities of neutron stars with those of hyperon stars.
The previously-found compensation mechanisms with microscopic Brueckner-Hartree-Fock calculations~\citep{2006PhRvC..73e8801S} are also observed here with a set of phenomenological EOS models, i.e., a stiffer nucleonic EOS will lead to an earlier onset of hyperons and thus a larger reduction of the maximum mass.
This is also the case for the statistical results as seen later in Sec.~\ref{sec:result} (from Table~\ref{tab:hyperonstar}). 

In a coalescing neutron star binary, changes in the orbital phasing due to the components’ mutual tidal interaction leave a detectable imprint in the gravitational wave signal, and the measured tidal deformabilities can then inform constraints on the neutron star EOS.
How easily the bulk matter in a star is deformed by an external tidal field is encoded in the tidal Love number $k_2$, the ratio of the induced quadruple moment $Q_{ij}$ to the applied tidal field $E_{ij}$~\citep{1992PhRvD..45.1017D,2009PhRvD..80h4035D,2008ApJ...677.1216H},
$Q_{ij}=-k_2\frac{2R^5}{3G}E_{ij}$.
The dimensionless tidal deformability $\Lambda$ is related to the compactness $M/R$ and the Love number $k_2$ through $\Lambda = \frac{2}{3}k_2(M/R)^{-5}$. 
The mass-weighed tidal deformability $\tilde{\Lambda}$ of a binary system,
\begin{eqnarray}
\tilde{\Lambda} = \frac{16}{13}\frac{(m_1+12m_2)m_1^4}{(m_1+m_2)^5}\Lambda_1 + (1\leftrightarrow2)\ ,
\end{eqnarray}
as a function of the chirp mass $\mathcal{M}=(m_1m_2)^{3/5}/(m_1+m_2)^{1/5}$, can be accurately measured during the inspiral.
In the presnt work, the tidal deformability measurement of the first binary neutron star merger event GW170817~\citep{2017PhRvL.119p1101A} is used to constrain the hyperon content in neutron stars and the underlying hyperonic interaction (see details below in Sec.~\ref{sec:analysis}). We also report in Sec.~\ref{sec:result} the most probable intervals of the tidal deformability for hyperon stars with typical mass of $1.4\Msun$.
Before presenting the statistical results, we introduce the Bayesian analysis framework in the next section. 
The statistical analysis is then preformed for 17 stiff EOSs (i.e., hereafter DD-PC1 is not included) that can support a $\geq2.3\Msun$ pure neutron star, ensuring that there is still room for the EOS softening by the addition of hyperons for explaining the measured masses of presently-heaviest $\sim2\Msun$ pulsars in binaries with white dwarfs~\citep{2010Natur.467.1081D,2013Sci...340..448A,2020NatAs...4...72C,2021ApJ...915L..12F}.

\section{Bayesian inference}  \label{sec:analysis} 

In Bayesian parameter estimation, the posterior distribution of a set of model parameters $\boldsymbol{\theta}$ given a dataset $\boldsymbol{D}$ is expressed as:
\begin{equation}
    P(\boldsymbol\theta|\boldsymbol{D}) = \frac{P(\boldsymbol{D}|\boldsymbol\theta)P(\boldsymbol\theta)}{\int P(\boldsymbol{D}|\boldsymbol\theta)P(\boldsymbol\theta){\rm d}\boldsymbol\theta}\ ,
\end{equation}
where $P(\boldsymbol\theta)$ is the prior probability of the parameter set $\boldsymbol\theta$. The total likelihood function $P(\boldsymbol{D}|\boldsymbol\theta)$ is given by the product of the likelihood $P_i({\boldsymbol d}_i|\boldsymbol\theta)$ of any individual observational data ${\boldsymbol d}_i\in\boldsymbol{D}$. 
In the following, we present a detailed discussion on the prior and likelihood we adopted in the analysis.

\begin{figure}
\includegraphics[width = 0.96\linewidth]{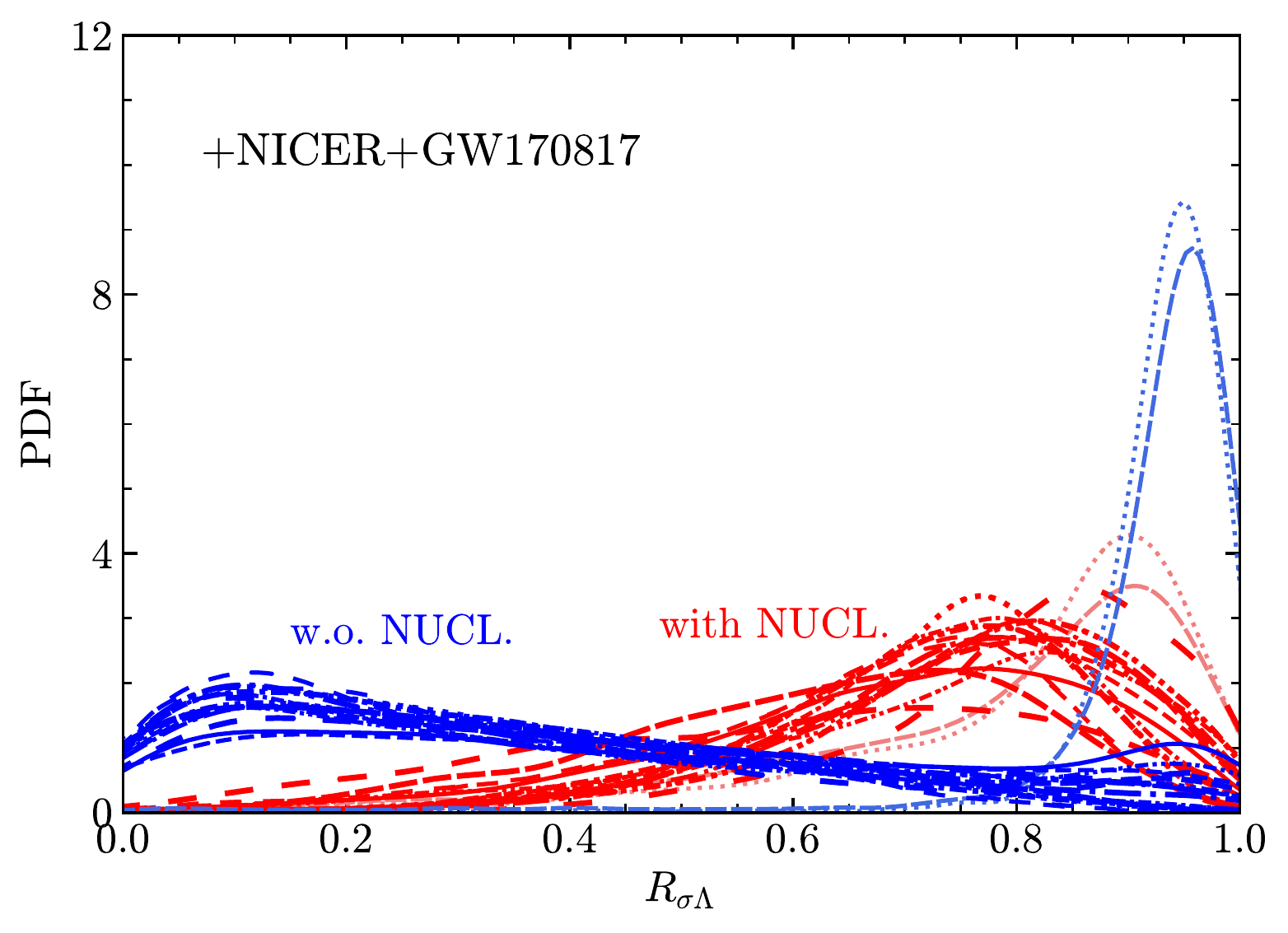}
\includegraphics[width = 0.96\linewidth]{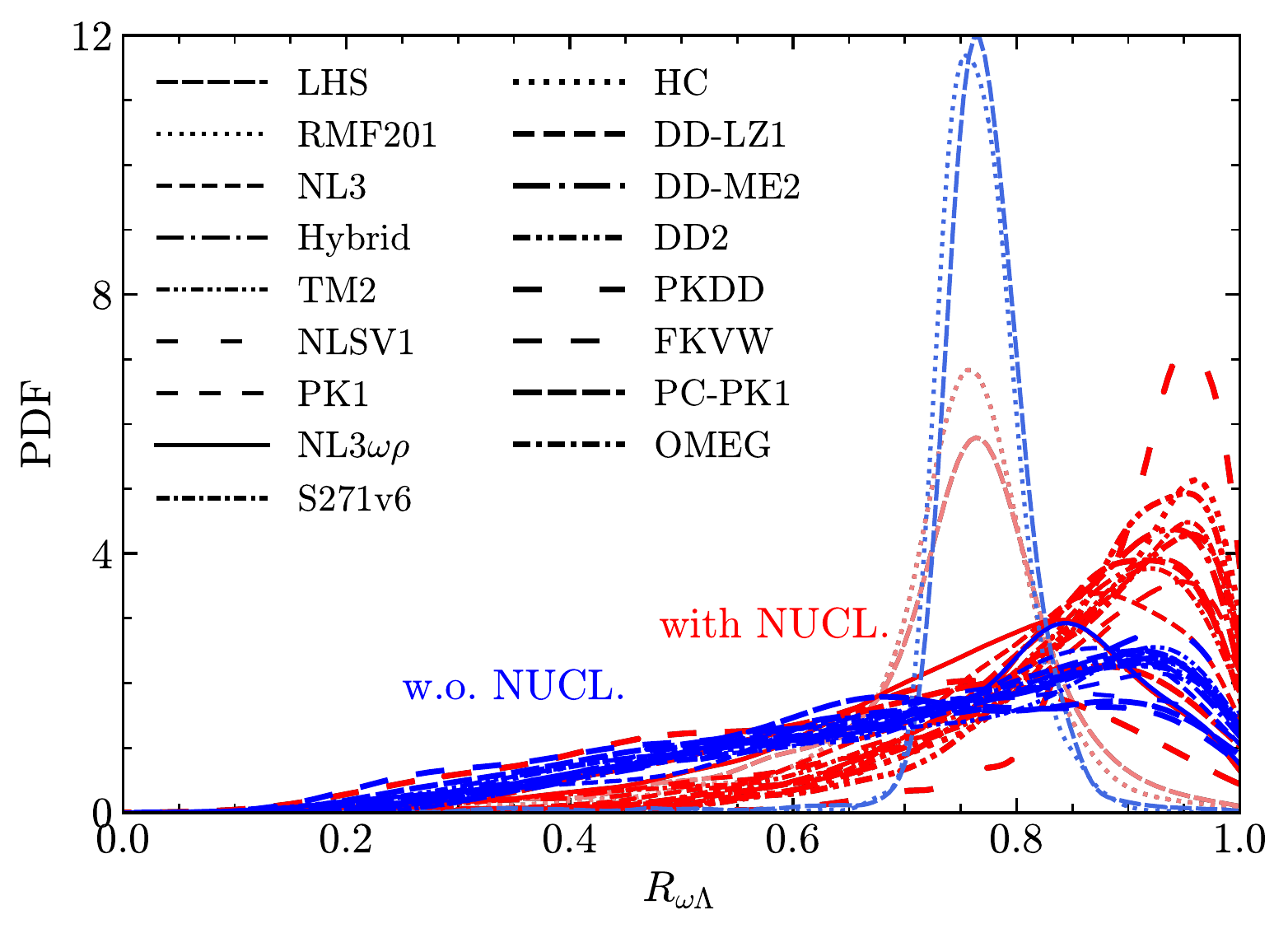} 	\vspace{-0.3cm}
\caption{Posterior PDFs of $R_{\sigma\Lambda}$ (upper panels) and $R_{\omega\Lambda}$ (lower panels) of scalar and vector couplings between $\Lambda N$ and $NN$ interactions for the considered relativistic EOSs. Both the data of GW170817 and NICER (PSR J0030+0451 and PSR J0740+6620) are incorporated, and the analysis are performed with (red curves) and without (blue curves) the inclusion of the empirical $R_{\sigma\Lambda}$-$R_{\omega\Lambda}$ relation constrained by available single $\Lambda$ hypernuclei. The results of the already-excluded LHS and RMF201 are shown in light colors (see Sec.~\ref{sec:eos2}).
} 	\vspace{-0.3cm}
\label{fig:6coupling}
\end{figure}
 
\begin{figure}
\includegraphics[width = 0.96\linewidth]{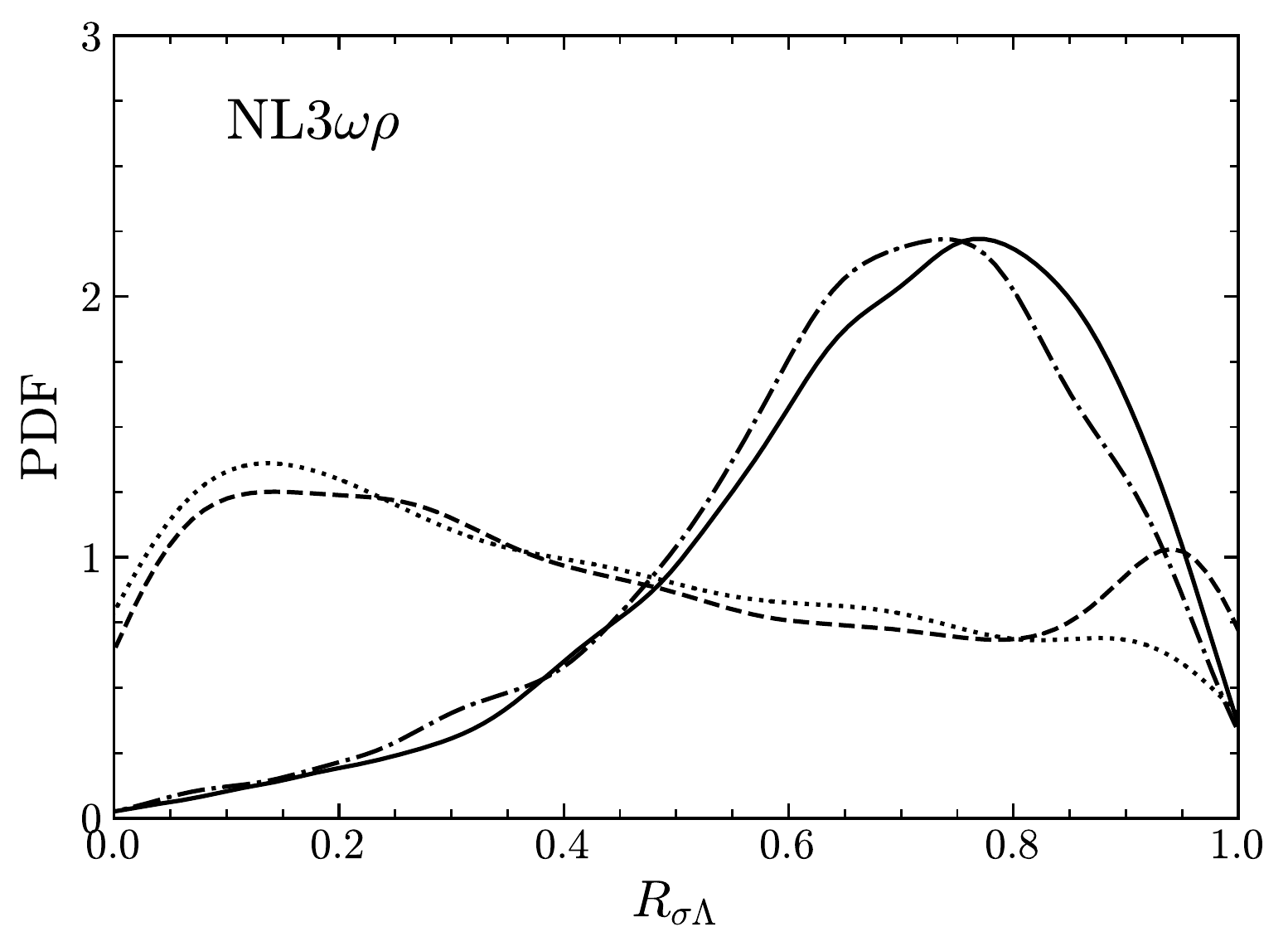}
\includegraphics[width = 0.96\linewidth]{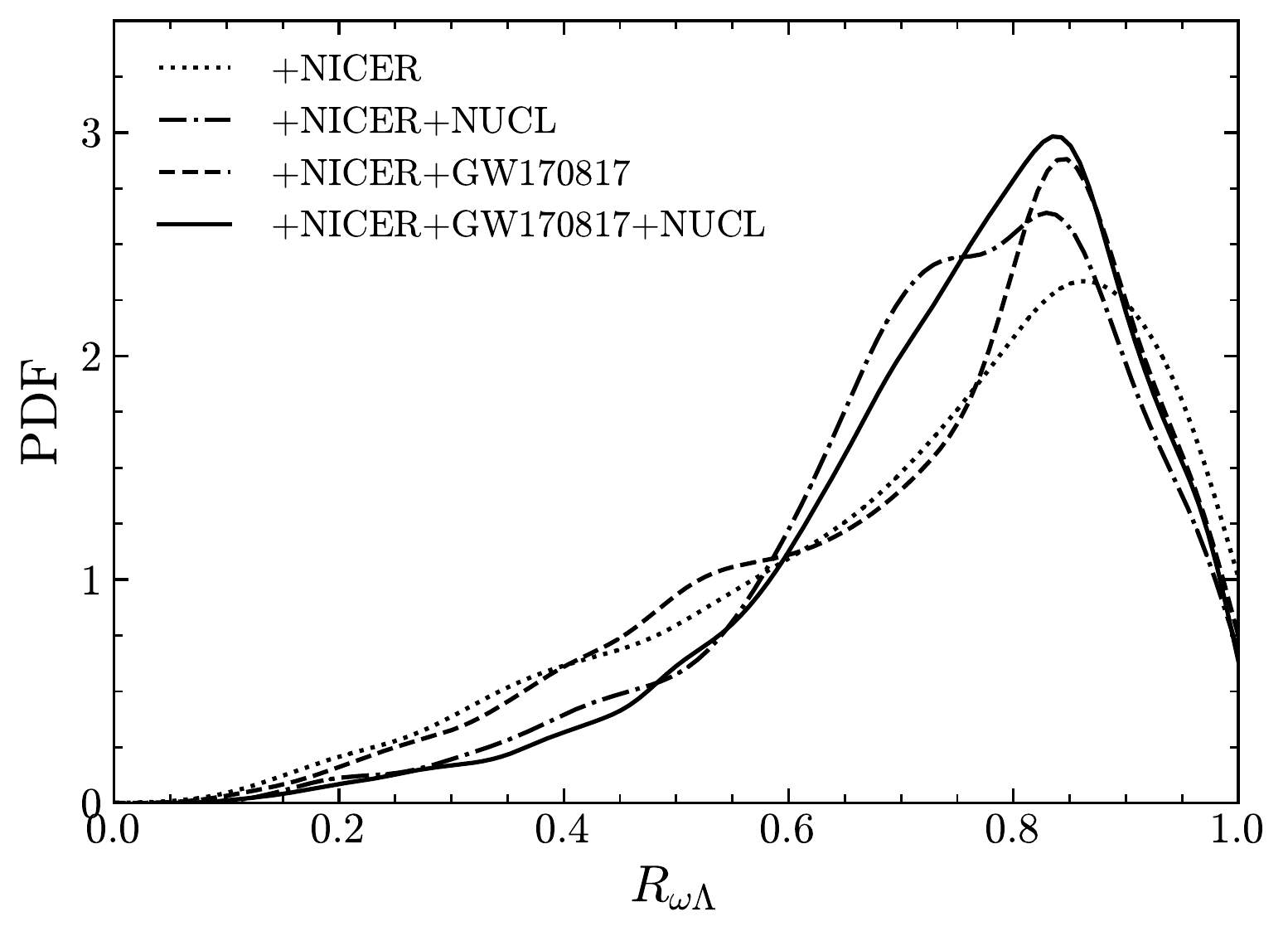} 	\vspace{-0.3cm}
\caption{Comparison of posterior PDFs of four different tests for $R_{\sigma\Lambda}$ (upper panels) and $R_{\omega\Lambda}$ (lower panels) detailed in Sec.~\ref{sec:analysis} using the NL3$\omega\rho$ EOS.
} 	\vspace{-0.3cm}
\label{fig:1coupling}
\end{figure}

\begin{figure}
\includegraphics[width = 0.96\linewidth]{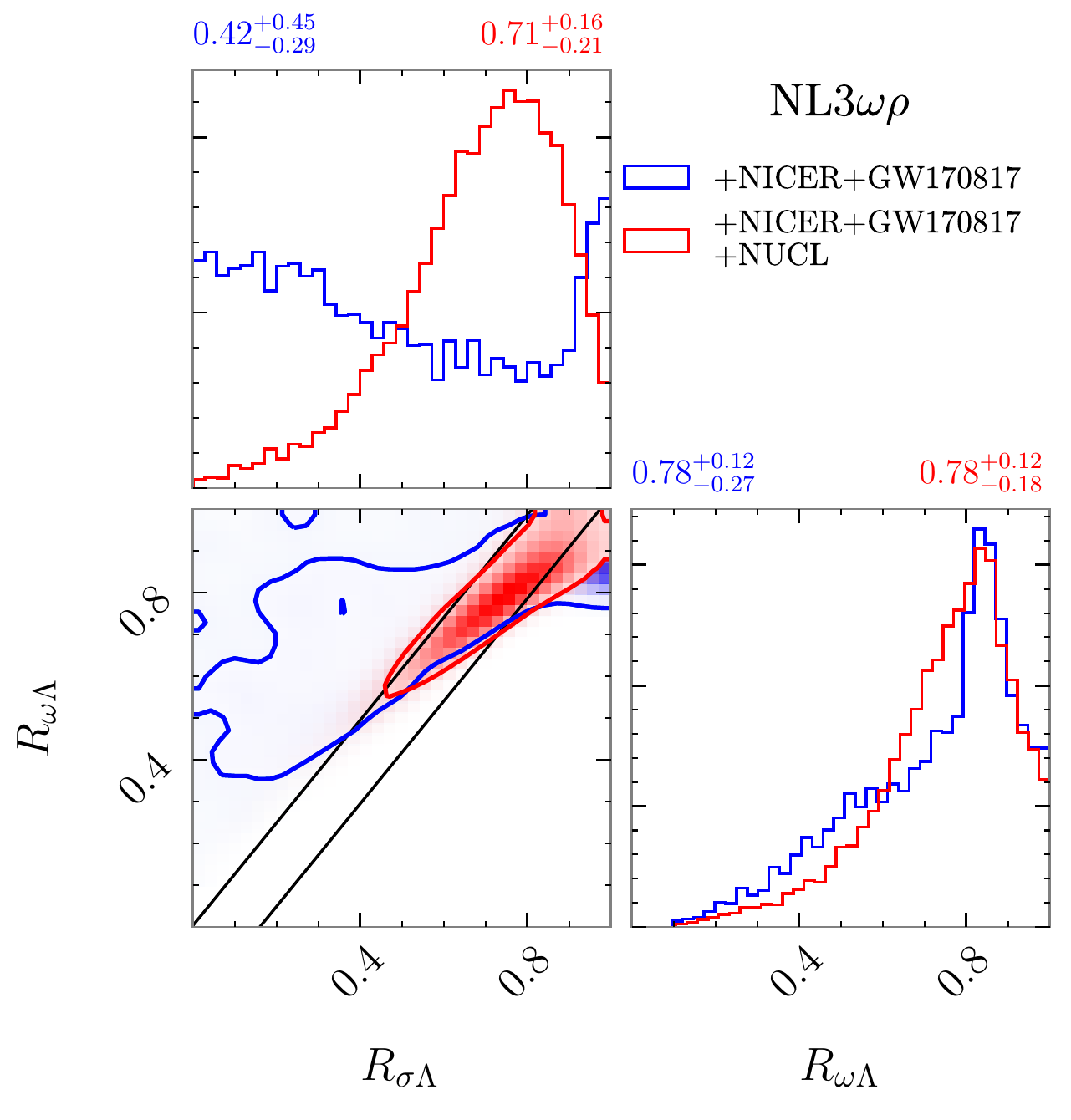} 	\vspace{-0.3cm}
\caption{Comparison of the posterior distributions of $R_{\sigma\Lambda}$ and $R_{\omega\Lambda}$ with (in red) and without (in blue) the empirical hypernuclei constraint.
The laboratory $R_{\sigma\Lambda}$-$R_{\omega\Lambda}$ relation, deduced from the measured $\Lambda$ separation energy in single $\Lambda$ hypernuclei, is also indicated with two black lines.
The contours are the $68\%$ credible regions for the parameters using the NL3$\omega\rho$ EOS. 
} 	\vspace{-0.3cm}
\label{fig:couplingcontour}
\end{figure}
 
\subsection{Dataset and likelihood}
In the present work, we will consider three types of experimental data: the mass-radius measurements of X-ray pulsars (PSR J0030+0451 and PSR J0740+6620) from NICER and the tidal deformability measurement (GW170817) from LIGO/Virgo, plus the laboratory hypernuclei data. 

\emph{NICER}. The NICER collaboration has reported two simultaneous mass-radius measurements of X-ray pulsars (PSR J0030+0451 and PSR J0740+6620) based on the pulse-profile modeling method. 
The results for PSR J0030+0451 were $M=1.34_{-0.16}^{+0.15}\Msun$, $R=12.71_{-1.19}^{+1.14}\,{\rm km}$~\citep{2019ApJ...887L..21R} or $M=1.44_{-0.14}^{+0.15}\Msun$, $R=13.02_{-1.06}^{+1.24}\,{\rm km}$~\citep{2019ApJ...887L..24M}. The results for PSR J0740+6620 were $M=2.072_{-0.066}^{+0.067}\Msun$, $R=12.39_{-0.98}^{+1.30}\,{\rm km}$~\citep{2021ApJ...918L..27R} or $M=2.062_{-0.091}^{+0.090}\Msun$, $R=13.71_{-1.50}^{+2.61}\,{\rm km}$~\citep{2021ApJ...918L..28M}. 
To incorporate the mass-radius data of these two sources, we take the likelihood function as:
\begin{equation}
    P_{\rm NICER}({\boldsymbol d}_{\rm NICER}|\boldsymbol\theta)=\prod_j P_j({M(\boldsymbol\theta),R(\boldsymbol\theta}))\ ,
\end{equation}
where we equate the individual likelihood $P_j$ to the joint posterior density distribution of $M$ and $R$ for PSR J0030+0451~\citep{2019ApJ...887L..21R} or for PSR J0740+6620~\citep{2021ApJ...918L..27R}.

\emph{GW170817}. The first detected double-neutron star merger is GW170817, with individual
masses between $1.17$ and $1.60\Msun$ with a total mass of $2.73_{-0.01}^{+0.04}\Msun$~\citep{2017PhRvL.119p1101A,2019PhRvX...9a1001A} and a chirp mass of $\mathcal{M}=1.186_{-0.001}^{+0.001}\Msun$~\citep{2019PhRvX...9a1001A}.
We calculate the likelihood of GW170817 through the interpolated likelihood table given by~\citet{2020MNRAS.499.5972H}, which is encapsulated in the python package \textsf{toast}~\footnote{\url{https://git.ligo.org/francisco.hernandez/toast}}. This interpolation table is obtained by fitting the strain data with the gravitational wave waveform from the component masses and their corresponding tidal deformabilities. The likelihood then reads:
\begin{equation}
    P_{\rm GW}({\boldsymbol d}_{\rm GW}|\boldsymbol\theta)= F(\Lambda_1(\boldsymbol\theta;M_1), \Lambda_2(\boldsymbol\theta;M_2), \mathcal M, q)\ ,
\end{equation}
where $F(\cdot)$ is the interpolation function. 
The component masses $M_1$ and $M_2$ are related to the chirp mass $\mathcal{M}$ and the mass ratio $q = M_2/M_1$.

\emph{NUCL}. Recently, from fitting calculated $\Lambda$ separation energies to experimental values of of eleven known $\Lambda$ hypernuclei with $A\ge12$, an excellent linear correlation between $R_{\sigma\Lambda}$ and $R_{\omega\Lambda}$ is found~\citep{2021PhRvC.104e4321R},  
i.e., $R_{\omega\Lambda}=1.228R_{\sigma\Lambda}-0.097$, and it is found that the relation is consistent with our chosen set of RMF models (see Appendix~\ref{A1} for details).
To encapsulate this empirical relation, we write the likelihood function as:
\begin{equation}
    P_{\rm NUCL}({\boldsymbol d}_{\rm NUCL}|\boldsymbol\theta)=\exp\left[-\frac12\frac{(R_{\sigma\Lambda}-\bar R_{\sigma\Lambda})^2}{\sigma^2_{R_{\sigma\Lambda}}}\right],
\end{equation}
where $\bar R_{\sigma\Lambda}=(R_{\omega\Lambda}+0.097)/1.228$ is deduced from the relation and $\sigma_{R_{\sigma\Lambda}}=0.08$ is taken to mimic the statistical error of $R_{\sigma\Lambda}$ in~\citet{2021PhRvC.104e4321R}. See Appendix~\ref{A2} for the results with different choices of the statistical error.

\subsection{Model parameters and priors}

In the present work, the model parameters can be divided into three groups:

1) The EOS parameters $\boldsymbol \theta_{\rm EOS} =\{R_{\sigma\Lambda},R_{\omega\Lambda}\}$. 
As discussed in the introduction, the hyperon-meson couplings should be weaker than the nucleon-meson couplings and not necessarily constrained by SU(3) symmetry.
We thus take uniform distributions for the coupling-strength ratios as $R_{\sigma\Lambda}\sim U[0,1]$ and $R_{\omega\Lambda}\sim U[0,1]$.

2) When we consider the NICER measurements, we add the central energy density of pulsar $j$, $\varepsilon_{c,j}$, into the parameter set to obtain its mass and radius, $M=M(\boldsymbol\theta_{\rm EOS};\varepsilon_{c,j})$ and $R=R(\boldsymbol\theta_{\rm EOS};\varepsilon_{c,j})$. 
We also assign reasonably wide ranges for the central energy density as $\varepsilon_{c}\sim U[0.6\times10^{15},3\times10^{15}]\,{\rm g/cm^3}$ for PSR J0030+0451, and as $\varepsilon_{c}\sim U[0.3\times10^{15},1\times10^{15}]\,{\rm g/cm^3}$ for PSR J0740+6620.

3) The gravitational wave parameters are the chirp mass $\mathcal{M}$ and the mass ratio $q$, while the tidal deformabilities of two components can be obtained through the EOS and component masses, $\Lambda_1(\boldsymbol\theta_{\rm EOS};M_1)$ and $\Lambda_2(\boldsymbol\theta_{\rm EOS};M_2)$. We take uniform priors for the chirp mass $\mathcal{M}\sim U[1.18,1.21]\,M_\odot$ and for the mass ratio $q\sim U[0.5,1]$.

With the priors and likelihood at hand, we then sample from the posterior distribution by using the python-based \textsf{bilby}~\citep{2019ascl.soft01011A} and \textsf{pymultinest}~\citep{2016ascl.soft06005B} packages.
We carry out four main tests to investigate the influence of individual astrophysical data as well as the laboratory data on the $\Lambda$-coupling strengths, namely:\\
(i)+NICER: where we consider both the constraints of PSR J0030+0451 and PSR J0740+6620 from NICER;\\
(ii)+NICER+NUCL: where we consider both the constraints in (i) and the hypernuclei one;\\
(iii)+NICER+GW170817: where we consider both the constraints in (i) and the GW170817 data;\\
(iv)+NICER+GW170817+NUCL: where we consider both the constraints in (iii) and the hypernuclei one.

\section{Results and discussion} \label{sec:result}

\begin{figure*}
         \includegraphics[width=2.3in]{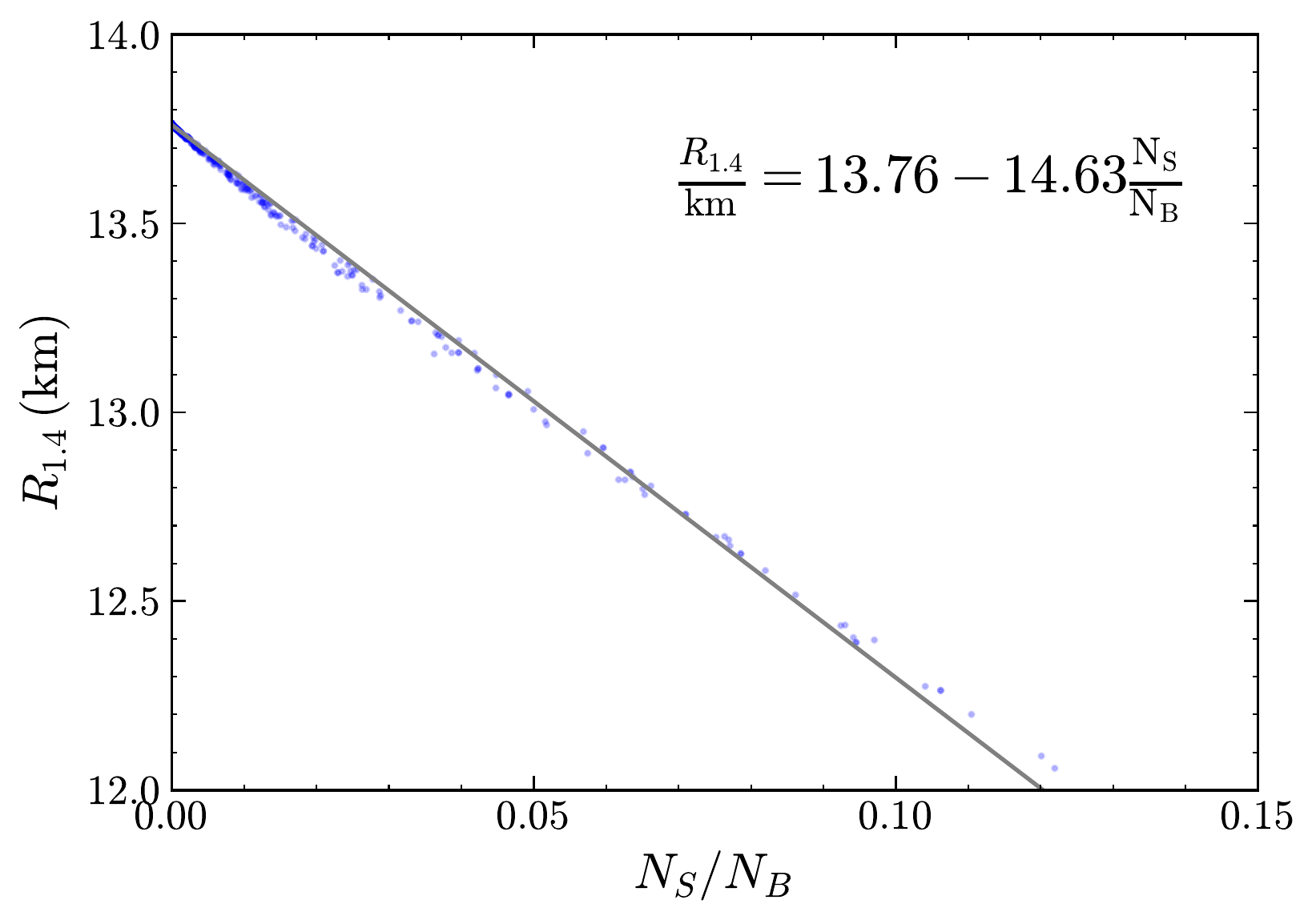}
         \includegraphics[width=2.3in]{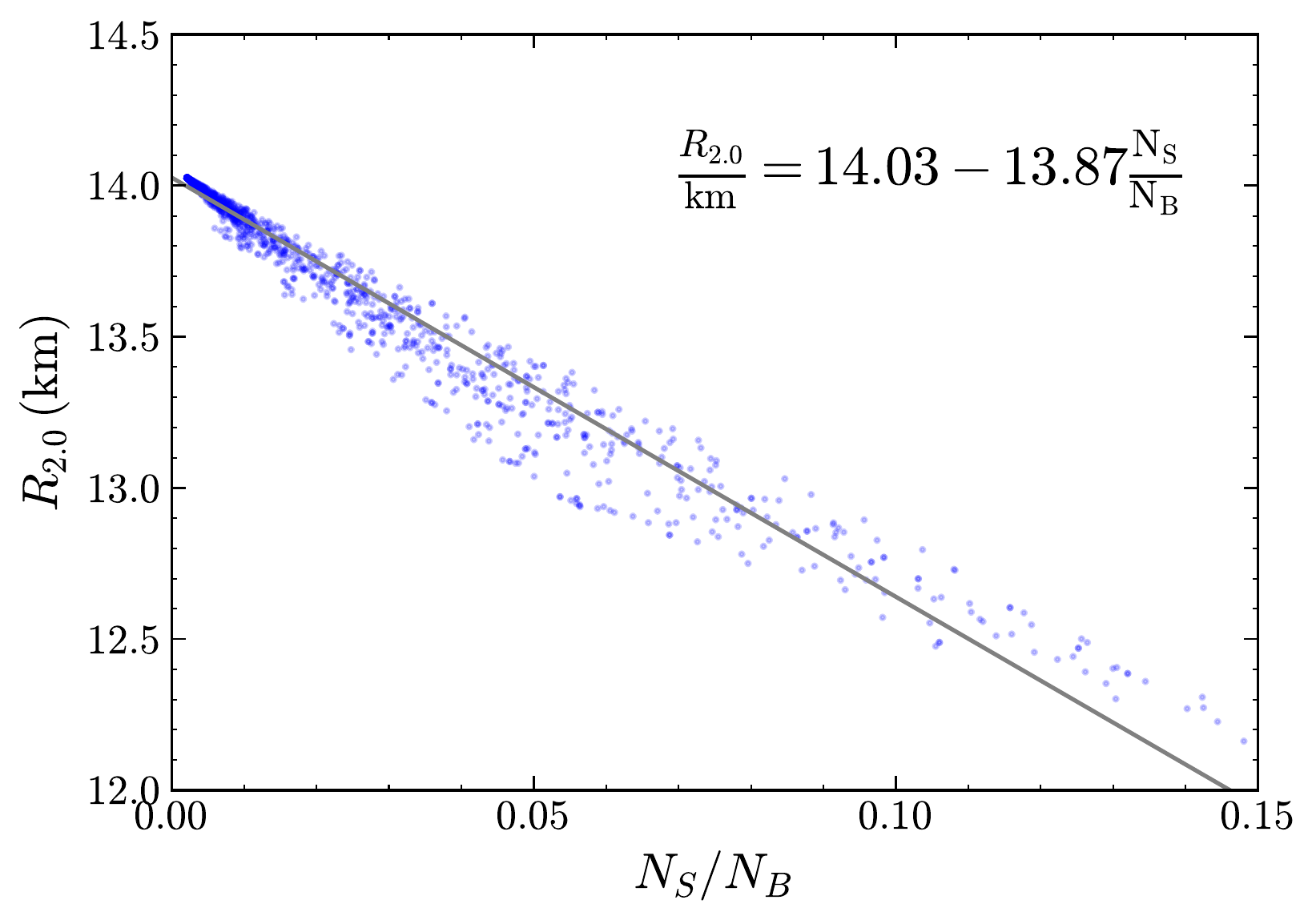}
         \includegraphics[width=2.3in]{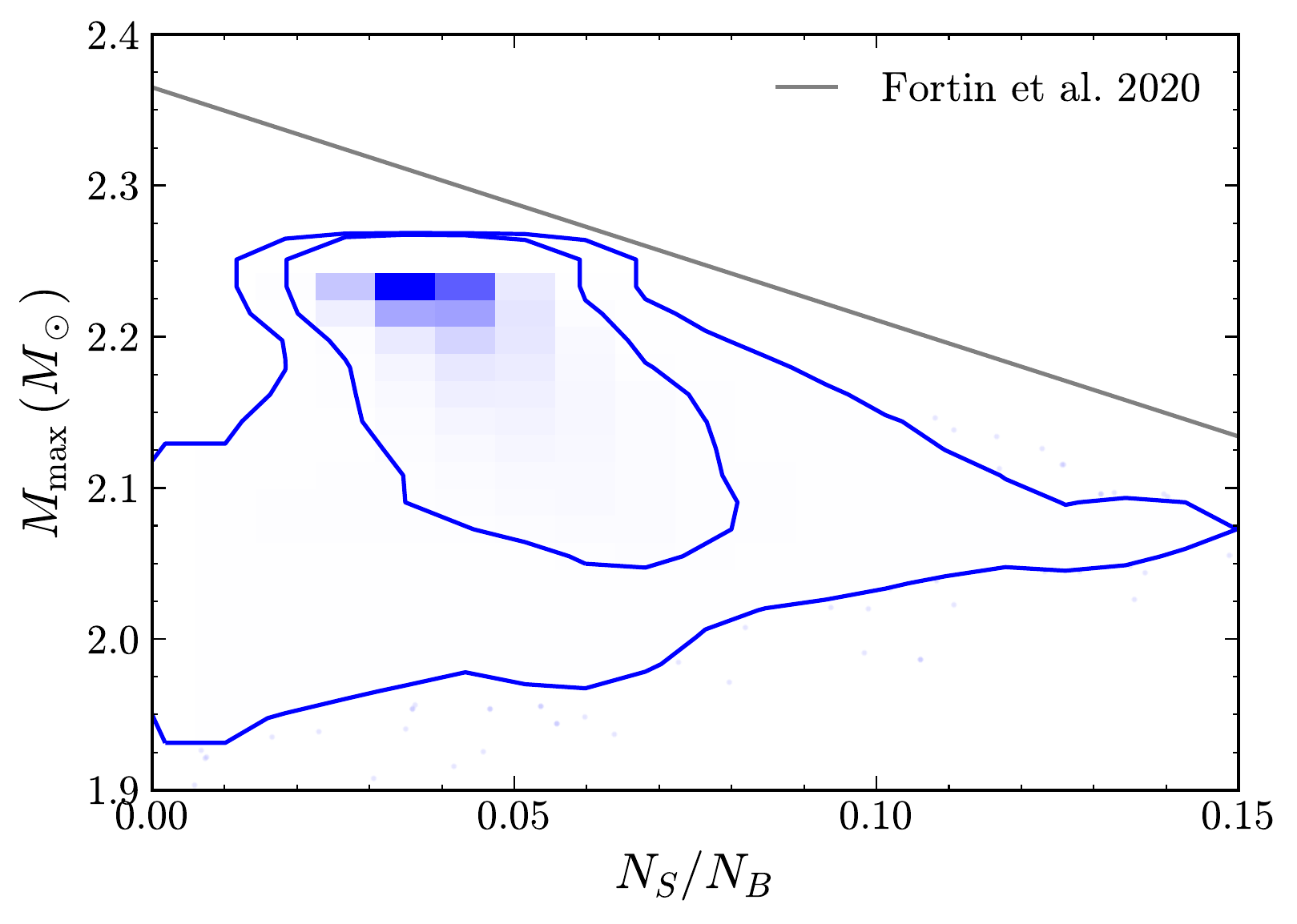} \vspace{-0.3cm}
\caption{Correlations between the strangeness fraction and the radius of $1.4\Msun$ stars (left panel), the radius of $2.0\Msun$ stars (middle panel), and the maximum mass of hyperon stars (right panel). The posterior distributions shown are conditioned on the joint analysis priors (+NICER+GW170817+NUCL). The left two panels are shown at $68\%$ credible level together with the newly-fitted linear functions, respectively, while both $68\%$ and $95\%$ credible regions are shown in the right panel for the maximum mass of hyperon stars with comparison to the previous obtained linear fit of \citet{2020PhRvD.101c4017F}. See text for details.
}	
\label{fig:fitting}
 \end{figure*}

\subsection{Hyperon-meson couplings}

The most probable values of the scalar and vector coupling ratios ($R_{\sigma\Lambda}$ and $R_{\omega\Lambda}$) and their $68\%$ credible boundaries, constrained by the astrophysical and laboratory data in four different tests, are reported in Table \ref{tab:coupling}. 
The posterior PDFs with the joint +NICER+GW170817 analysis on all 17 EOSs are shown in Fig.~\ref{fig:6coupling}.
Two additional figures are reported for the representative NL3$\omega\rho$ case: Fig.~\ref{fig:1coupling}
for showing the effects of including or excluding any individual data, and Fig.~\ref{fig:couplingcontour} for comparing the $R_{\sigma\Lambda}$ and $R_{\omega\Lambda}$ results from the astrophysical and laboratory data.

From Fig.~\ref{fig:6coupling}, expect the already-excluded LHS~\citep{1989RPPh...52..439R} and RMF201~\citep{2010PhRvC..82b5203D}, it is evident that the introduction of the hypernuclei constraint favors large values of $R_{\sigma\Lambda}$ and $R_{\omega\Lambda}$ and disfavors small values of both couplings: The peak values of $R_{\sigma\Lambda}$ are shifted largely to the right, while the $R_{\omega\Lambda}$ peaks only slightly change; Nevertheless, the $R_{\omega\Lambda}$ distributions are considerably narrowed towards to the right.
The strong underlying $R_{\sigma\Lambda}$-$R_{\omega\Lambda}$ correlation imposed by the data of single $\Lambda$ hypernuclei ensures a large enough scalar hyperon coupling to match the large vector hyperon coupling. For example, in the case of NL3$\omega\rho$, $R_{\sigma\Lambda}$ is lifted from $\sim0.4$ to $\sim0.7$ when the hypernuclei constraint is added.
Also, the softer the EOS, the more concentrated the distributions for large values of $R_{\sigma\Lambda}$ and $R_{\omega\Lambda}$.
One may notice the second peaks at large $R_{\omega\Lambda}$ values for NL3$\omega\rho$ and PKDD are removed by the introduction of the hypernuclei constraint.
As seen more clearly in the NL3$\omega\rho$ plot of Fig.~\ref{fig:1coupling}, such peaks are present because the not-so-high tidal deformability upper limit of GW170817 necessarily requires a considerable scalar attraction between nucleons and hyperons.
The addtion of astrophysical observational data on top of the laboratory  $R_{\sigma\Lambda}$-$R_{\omega\Lambda}$ correlation also rotates the linear correlation (indicated with two black lines in Fig.~\ref{fig:couplingcontour} slightly towards the direction of small values of $R_{\omega\Lambda}$.
One sees the exclusion of the small-$R_{\sigma\Lambda}$ area in the upper-left corner of Fig.~\ref{fig:couplingcontour} by the hypernuclei constraint. 

\subsection{Hyperon star properties}
 
\begin{table*}
	\centering
	\caption{Most probable intervals of various hyperon star properties for 17 RMF effective interactions, to the $68\%$ confidence level, constrained jointly by the +NICER+GW170817+NUCL analysis explained in Sec.~\ref{sec:analysis}.
	The corresponding results of neutron stars (without hyperons) are also shown.
	  $M_{\rm max}$ is the maximum mass and $\rho_c/\rho_0$ is the corresponding central density scaled by the saturation density. $R_{2.0}$ is the radius of $2.0\Msun$ stars. $R_{1.4}$ and $\Lambda_{1.4}$ are the radius and tidal deformability of $1.4\Msun$ stars, respectively. The $R_{2.0}$ results for eight relatively soft EOSs (NLSV1, PK1, S271v6, HC, DD2, PKDD, PC-PK1, OMEG) are not shown because most of the posterior hyperon stars can not reach $2.0\Msun$.	  
	  }	\label{tab:hyperonstar}
	\vspace{-0.3cm}
\renewcommand\arraystretch{1.2}
  \begin{ruledtabular}
	\begin{tabular}{cccccccc} 
			EOS & $ $ &$M_{\rm max}/{\rm M_\odot}$ &$\rho_c/\rho_0$ &$R_{2.0}/{\rm km}$ &$R_{1.4}/{\rm km}$ & $\Lambda_{1.4}$     \\
			\hline
			\multirow{2}{*}{LHS}
			&w.o. $Y$ &$2.986$&$3.919$ &$15.449$&$15.083$ & $1643.781$  \\
			&with $Y$&$2.106^{+0.335}_{-0.119}$&$4.628^{+0.901}_{-0.319}$&$13.190^{+2.135}_{-0.706}$&$14.049^{+0.000}_{-0.947}$&$1643.781^{+0.000}_{-630.280}$  \\
			\hline
			\multirow{2}{*}{RMF201}
			&w.o. $Y$ &$2.953$&$3.882$ &$15.212$&$14.810$ & $1526.412$  \\
			&with $Y$&$2.117^{+0.307}_{-0.058}$&$4.556^{+1.293}_{-0.058}$&$13.486^{+1.335}_{-0.533}$&$14.369^{+0.000}_{-0.331}$&$1524.401^{+0.000}_{-622.201}$  \\
			\hline
			\multirow{2}{*}{NL3}
			&w.o. $Y$ &$2.777$&$4.514$ &$14.784$&$14.827$ & $1258.012$  \\
			&with $Y$&$2.147^{+0.041}_{-0.214}$&$4.541^{+1.351}_{-0.205}$&$14.676^{+0.335}_{-1.326}$&$14.824^{+0.000}_{-0.768}$&$1258.012^{+0.000}_{-146.888}$  \\
			\hline
			\multirow{2}{*}{Hybrid}
			&w.o. $Y$ &$2.773$&$4.554$ &$14.683$&$14.758$ & $1161.133$  \\
			&with $Y$&$2.131^{+0.034}_{-0.209}$&$4.649^{+1.106}_{-0.213}$&$14.523^{+0.468}_{-1.350}$&$14.588^{+0.000}_{-0.205}$&$1161.131^{+0.000}_{-25.890}$  \\
			\hline
			\multirow{2}{*}{FKVW}
			&w.o. $Y$  &$2.747$&$4.353$&$14.360$&$14.121$ & $1106.960$ \\
			&with $Y$ &$2.164^{+0.065}_{-0.190}$&$3.758^{+0.303}_{—0.132}$&$14.288^{+0.085}_{-1.132}$&$14.079^{+0.000}_{-0.006}$&$1092.946^{+0.000}_{-9.728}$\\
			\hline			
			\multirow{2}{*}{NL3$\omega\rho$}
			&w.o. $Y$ &$2.756$&$4.676$ &$14.070$&$13.772$ & $941.852$  \\
			&with $Y$&$2.176^{+0.085}_{-0.202}$&$4.846^{+0.046}_{-0.501}$&$13.968^{+0.096}_{-1.512}$&$13.769^{+0.000}_{-1.084}$&$940.165^{+0.000}_{-443.756}$  \\
			\hline
			\multirow{2}{*}{DDLZ1}
			&w.o. $Y$ &$2.560$&$4.942$&$13.376$ &$13.146$&$727.072$  \\
			&with $Y$ &$2.065^{+0.009}_{-0.076}$&$5.120^{+0.025}_{-0.044}$&$12.976^{+0.027}_{-0.230}$&$13.141^{+0.000}_{-0.581}$&$725.192^{+0.000}_{-279.012}$  \\
			\hline			
			\multirow{2}{*}{TM2}
			&w.o. $Y$ &$2.492$&$5.143$ &$15.277$&$15.460$ & $1677.406$  \\
			&with $Y$&$2.089^{+0.025}_{-0.218}$&$4.470^{+1.379}_{-0.223}$&$15.237^{+0.004}_{-0.635}$&$15.460^{+0.000}_{-0.162}$&$1677.402^{+0.000}_{-104.596}$  \\
			\hline
			\multirow{2}{*}{DD-ME2}
			&w.o. $Y$ &$2.486$&$5.319$&$13.286$&$13.241$&$730.747$    \\
			&with $Y$ &$2.020^{+0.003}_{-0.049}$&$5.539^{+0.178}_{-0.006}$&$12.654^{+0.019}_{-0.215}$&$13.238^{+0.000}_{-0.270}$ &$711.558^{+0.000}_{-168.432}$\\
			\hline
			\multirow{2}{*}{OMEG}
			&w.o. $Y$  &$2.457$&$5.318$&$12.882$&$12.978$ & $576.307$ \\
			&with $Y$ &$1.973^{+0.010}_{-0.111}$&$5.381^{+0.145}_{—0.008}$& -&$12.977^{+0.000}_{-0.033}$&$565.429^{+0.000}_{-2.297}$\\
			\hline
			\multirow{2}{*}{DD2}
			&w.o. $Y$  &$2.419$&$5.282$&$13.133$&$13.212$      &$ 639.034$ \\
			&with $Y$ &$1.945^{+0.004}_{-0.010}$&$5.913^{+0.154}_{-0.007}$& -&$13.210^{+0.000}_{-0.398}$ &$635.582^{+0.000}_{-110.543}$  \\
			\hline
			\multirow{2}{*}{S271v6}
			&w.o. $Y$ &$2.35$&$6.013$ &$12.889$&$13.167$ & $635.925$  \\
			&with $Y$&$1.927^{+0.045}_{-0.095}$&$7.342^{+0.932}_{-0.301}$& -&$13.166^{+0.000}_{-0.019}$& $635.921^{+0.000}_{-8.544}$  \\	
			\hline
			\multirow{2}{*}{NLSV1}
			&w.o. $Y$ &$2.344$&$5.142$ &$14.053$&$14.495$ & $1028.908$  \\
			&with $Y$&$1.886^{+0.041}_{-0.051}$&$4.322^{+0.107}_{-0.044}$& -&$14.495^{+0.000}_{-0.026}$&$1028.901^{+0.000}_{-1.063}$  \\
			\hline	
			\multirow{2}{*}{PKDD}
			&w.o. $Y$ &$2.329$&$5.920$ &$13.223$&$13.710$ & $768.771$   \\
			&with $Y$ &$1.818^{+0.005}_{-0.037}$&$6.170^{+0.323}_{-0.063}$& -&$13.709^{+0.000}_{-1.257}$ &$764.093^{+0.000}_{-394.699}$    \\		
			\hline
			\multirow{2}{*}{PK1}
			&w.o. $Y$  &$2.315$&$5.348$&$14.058$&$14.531$ & $1111.863$ \\
			&with $Y$ &$1.912^{+0.000}_{-0.072}$&$4.696^{+0.378}_{-0.034}$& -&$14.529^{+0.000}_{-0.003}$&$1102.674^{+0.000}_{-3.318}$\\
			\hline			
			\multirow{2}{*}{PC-PK1}
			&w.o. $Y$  &$2.306$&$4.422$&$14.188$&$14.442$ & $1079.255$ \\
			&with $Y$ &$1.852^{+0.065}_{-0.190}$&$5.198^{+0.272}_{—0.022}$& -&$14.421^{+0.000}_{-0.064}$&$1079.249^{+0.000}_{-35.556}$\\			
			\hline
			\multirow{2}{*}{HC}
			&w.o. $Y$ &$2.286$&$5.775$ &$12.141$&$12.410$ & $446.984$  \\
			&with $Y$&$1.828^{+0.010}_{-0.051}$&$7.423^{+0.053}_{-0.005}$& -&$12.410^{+0.000}_{-0.000}$&$446.984^{+0.000}_{-0.000}$&  \\
	\end{tabular}
\end{ruledtabular}
\end{table*}

Table \ref{tab:hyperonstar} collects various hyperon star properties for all 17 considered EOSs, at $68\%$ credible level, under the most probable hyperonic interaction from the joint +NICER+GW170817+NUCL analysis. 
For the representative stiff NL3$\omega\rho$ EOS case, we further present in Fig.~\ref{fig:hyperonstar} the composition and the mass versus radius \& density relations of hyperon stars in comparison with the corresponding neutrons star results. 
Since the NL3$\omega\rho$ EOS is one the stiffest ones in the literature, while being consistent with all current nuclear and astrophysical constraints, from the figure, one may learn roughly the status of uncertainties of the hypernuclear matter and hyperon star properties due to the uncertain hyperon-nucleon interaction in dense matter. 

In the left panel of Fig.~\ref{fig:hyperonstar} for the particle composition, we see that at densities below saturation, the charge-neutral matter is almost pure neutrons with a small admixture of protons and
electrons; With increasing density, the electron Fermi energy rises to the muon mass, and the muons will be populated. Hyperon thresholds are reached at above $\sim1.5\rho_0$, and the negatively charged $e^-$ and $\mu^-$ are then replaced by negatively charged hyperons with the increase of density.
At densities close to the central density of the maximum-mass hyperon stars ($\sim5\rho_0$; See Table \ref{tab:hyperonstar}), the hyperons constitute a sizable fraction of baryons, and the stars are essentially baryonic stars without the lepton population.
Nevertheless, the lack of precise knowledge of hyperonic interaction strongly affects the predicted hyperonic composition of neutron-star cores; It is still unclear whether the lightest $\Lambda$ of the baryon octet or the heavier but negatively-charged $\Xi^-$ appear first. And the threshold density of $\Lambda$ hyperons is in the wide range of $\sim1.4$-$3.8\rho_0$. 
In the current choice of the strength of hyperonic interaction, three hyperon species ($\Lambda$, $\Xi^-$, $\Xi^0$) are present; $\Sigma$ hyperons are absent because they interact repulsively in the dense medium.
Despite a strong dependence on the uncertain hyperonic interaction, one can notice generic features that the formation of hyperons softens the EOS.
This is because the Fermi pressure of neutrons and protons near the top of their Fermi seas is relieved by allowing them to hyperonize to unoccupied low-momentum states, producing lower pressure.
Consequently, neutron stars containing hyperons are more compact, and the maximum mass is lowered by about $20\%$, for example, from $2.7\Msun$ down to $\sim2.2\Msun$ in the NL3$\omega\rho$ case.
And a stiffer nucleon EOS leads to more softening (thus more decreasing in the maximum mass) since the hyperon appearance shifts to lower densities (see Table \ref{tab:hyperonstar} in the 3rd column from the left).
As to the smaller radius due to hyperons~\citep[e.g.,][]{2012PhRvC..85b5806L}, the effect sets in above $\sim0.5\Msun$ and grows with the stellar mass: $\sim16\%$ for $2.0\Msun$ stars and relatively half for $1.4\Msun$ stars, as seen in the right panel of Fig.~\ref{fig:hyperonstar}. 
We mention here that, according to the results of the representative stiff NL3$\omega\rho$ EOS, at $68\%$ credible level, the maximum of hyperon star is $M_{\rm max}=2.176^{+0.085}_{-0.202}\Msun$, with the peak value being $\sim2.26\Msun$ with the current determination of hyperonic interaction. Therefore it is not supported that the secondary components of $\rm GW190814$ (with mass $\sim2.6\Msun$)~\citep{2020ApJ...896L..44A} and $\rm GW200210\_092254$ (with mass $\sim2.8\Msun$)~\citep{2021arXiv211103606T} are hyperon stars.

Before the end of the section, we make an attempt to provide possibly useful relations between observed properties and the underlying phase state of hyperons stars based on our analysis.
The strangeness fraction $N_S/N_B$, defined as follows, 
\begin{equation}
\begin{split}
    N_S = \frac{4\pi}{3}\int dr\frac{q_{Si}\rho_ir^2}{\sqrt{1-2m(r)/r}}\ ;\\
    N_B = {4\pi}\int dr\frac{\rho_ir^2}{\sqrt{1-2m(r)/r}}\ ,    
\end{split}
\end{equation}
characterizes the strangeness contents in hypernuclear matter, $\rho_i$ and $q_{Si}$ representing the particle number density and strangeness charge of particle $i$, respectively. 
$m(r)$ denotes the enclosed gravitational mass at radial coordinate $r$. 
Fig.~\ref{fig:fitting} shows the posterior distributions of $R_{\rm 1.4}$, $R_{\rm 2.0}$ and $M_{\rm max}$ as functions of the strangeness fraction $N_S/N_B$ from the joint analysis of our representative EOS.
There are excellent linear anti-correlations between $N_S/N_B$ and $R_{\rm 1.4}$ as well as $R_{\rm 2.0}$ as shown in the left two panels of Fig.~\ref{fig:fitting}, with the determination coefficients of $R^2=0.998$ and $R^2=0.968$, respectively. 
There appear also an anti-correlation between $N_S/N_B$ and $M_{\rm max}$ of hyperon star, and the tread is similar with what previously revealed in~\citet{2020PhRvD.101c4017F}, but no simple relation is found in the present analysis.

\section{Summary}

The dense neutron-star matter may contain a significant fraction of non-nucleon baryonic components, such as strangeness-bearing hyperons. They affect the stellar structure and evolution in various ways, see e.g.~\citet{2021PrPNP.12003879B,2021Univ....7..408L} for recent reviews. 
The formation of hyperons inside neutron stars may lead to a significant softening of the EOS of a neutron-star core, with respect to the $npe\mu$ case, and lower the theoretical maximum mass below the observed pulsar masses, causing the well-known hyperon puzzle problem~\citep[e.g.,][]{2011PhRvC..83b5804B,2015PhRvL.114i2301L,2017hspp.confj1002B}. 
The hyperon softening of the EOS may strongly affect the spin evolution of isolated neutron stars~\citep[e.g.,][]{2004A&A...416.1013Z,2015ChPhL..32k2101Q}; 
And sufficiently massive stars containing hyperons (or other negatively-charged, strongly-interacting particles) may result in a delayed collapse of the hot newly-born neutron star to a black hole~\citep{1996ApJ...468..823B}.
The presence of hyperons may affect the cooling of neutron stars~\citep[e.g.,][]{2009ApJ...691..621T}, the dynamics and the gravitational wave radiation of neutron star mergers~\citep[e.g.,][]{2021PhRvC.103c5810P,2021PhRvC.104b5801B}.
Also, as soon as hyperons appear, various hyperon reactions contribute to the bulk viscosity of neutron star matter and influence the damping of r-mode instability of neutron stars~\citep[e.g.,][]{1969Ap&SS...5..213L}. 
The resolving of these problems necessarily requires an accurate description of the hyperonic interaction.

In the present work, we performed one of the first Bayesian inferences of the hyperon-nucleon interaction strengths in the relativistic Lagrangian using the robust multimessenger neutron star observations from LIGO/Virgo and NICER, in combination with the study of hypernuclei. 
A set of 17 RMF interactions (LHS, RMF201, NL3, Hybrid, TM2, NLSV1, PK1, NL3$\omega\rho$, S271v6, HC, DD-LZ1, DD-ME2, DD2, PKDD, PK1, FKVW, PC-PK1 and OMEG).
The medium dependence of the effective mean-field interactions, accounting for higher-order many-body effects, is also included. 
The accurately-measured mass, radius (for PSR J0030+0451 and PSR J0740+6620) and tidal deformability (for GW170817) of neutron stars can be translated directly into the information on the underlying EOS including hyperons. 
We mainly focus on the possible constraints on $\Lambda$ couplings in dense stellar medium, by confronting the single $\Lambda$-hypernuclei data with the neutron star observational data.
In particular, we relax the SU(3) symmetry commonly assumed and take the priors distributions of the coupling ratios $R_{\sigma \Lambda}=g_{\sigma \Lambda}/g_{\sigma N}$, $R_{\omega \Lambda}=g_{\omega \Lambda}/g_{\omega N}$ as uniform ones in the ranges of 0 to 1.
The Bayesian analysis is performed for four different tests with or without the laboratory hypernuclear constraint, and $R_{\sigma \Lambda}$, $R_{\omega \Lambda}$ are studied by mapping the hyperon-star parameter space with the observational data.  

The laboratory constraint from the single $\Lambda$-hypernuclei data is found to play an important role in determining the phenomenological interactions, preventing a too-small scalar coupling through a strong positive correlation between the scalar and vector ones.
Also, consistent with previous studies, a stiff (soft) nucleon EOS causes an
earlier (later) onset and larger (smaller) concentration of hyperons in neutron star interior, which considerably change the star properties.
Among the set of employed EOSs, we take the NL3$\omega\rho$ one as an exemplary stiffest one and discuss some quantitative features for hyperon stars in the light of the statistical results with NL3$\omega\rho$.  
We find the hyperon threshold can be as low as $\sim1.5$ times the nuclear saturation density, and the hyperon star EOS is moderately stiff with the $68\%$ credible interval of the maximum mass being $M_{\rm max}=2.176^{+0.085}_{-0.202}\Msun$.
The introduction of hyperons not only softens the neutron star EOS, but also results in a more compact star. And we find the decrease of the radii of typical $1.4\Msun$ and $2.0\Msun$ stars can both be linearly depicted as functions of the strangeness fraction in neutron stars. 
The corresponding analytic fittings are provided, potentially useful for future research on the structure and evolution of hyperon stars. 

There are several caveats in the present work. 
The major one is that, for reasons of simplicity, we only explore the preferred coupling constants of $\Lambda$ hyperons, while keeping the $\Sigma$ and $\Xi$ hyperon couplings fixed to their empirical values. 
We take the simplified procedure for the study also because similar studies for $\Sigma$ and $\Xi$ hyperon couplings should await considerable progress on $\Sigma$ and $\Xi$ hypernuclei in the future.
Recently, the possibility of hyperons in equilibrium with other exotic phases [such as kaons~\citep[e.g.,][]{1995PhLB..349...11E,2007ChPhy..16.1934L}, quarks~\citep[e.g.,][]{2007PhRvD..76l3015M,2016EPJA...52...65M}, $\Delta$-resonances~\citep[e.g.,][]{2018PhLB..783..234L,2019ApJ...883..168R}] becomes a vivid field of research.
For future plans, it is interesting to consider multiple forms of exotic phases and study their competition and coexistence.
In addition, a further RMF model is an effective Lagrangian with meson fields mediating strong interactions between quarks, which we call the quark-meson coupling model~\citep{1988PhLB..200..235G}, or the quark mean-field model~\citep{2000PhRvC..61d5205S,1998PhRvC..58.3749T}.
It self-consistently relates the internal quark structure of a nucleon and a hyperon to the RMFs arising in the nuclear and hypernuclear matter, respectively, and has been employed extensively in the calculations of finite (hyperon-)nuclei and infinite dense matter. See~\citet{2018PrPNP.100..262G,2020JHEAp..28...19L} for recent reviews.
Therefore the present study can be further extended in this direction.

\appendix
\section{Linear relations between $R_{\sigma\Lambda}$ and $R_{\omega\Lambda}$ with different RMF models}\label{A1}

As suggested in~\citet{2021PhRvC.104e4321R}, the linear relation between $R_{\sigma\Lambda}$ and $R_{\omega\Lambda}$ for finite-range RMF effective interactions can be expressed as:
\begin{equation}\label{eq:RMF}
    R_{\omega\Lambda} \approx \frac{-g_{\sigma N}\sigma}{U_{N} 
    -g_{\sigma N}\sigma}R_{\sigma \Lambda}
    +\frac{U_{\Lambda}}{U_{N} -g_{\sigma N}\sigma}\ ,
\end{equation}
while for zero-range PC models, an analogous relation reads~\citep{2012PhRvC..85a4306T}:
\begin{equation}\label{eq:PCRMF}
\begin{aligned}
    R_{\omega\Lambda} &\approx 
    \frac{-\alpha_S^{NN}\rho_S }{U_{N} - \alpha_S^{NN}\rho_S -\beta_S^{NN}\rho_S^2-\gamma_S^{NN}\rho_S^3 - \gamma_V^{NN}\rho_{V}^3
    -\alpha_S^{N\Lambda}\rho_S^{\Lambda}-\alpha_V^{N\Lambda}\rho_V^{\Lambda}}R_{\sigma\Lambda}\\\nonumber
    & + \frac{U_{\Lambda}}{U_{N} - \alpha_S^{NN}\rho_S -\beta_S^{NN}\rho_S^2-\gamma_S^{NN}\rho_S^3 - \gamma_V^{NN}\rho_{V}^3
    -\alpha_S^{N\Lambda}\rho_S^{\Lambda}-\alpha_V^{N\Lambda}\rho_V^{\Lambda}}\ ,
\end{aligned}
\end{equation}

\begin{figure}
\centering
\includegraphics[width = 0.56\linewidth]{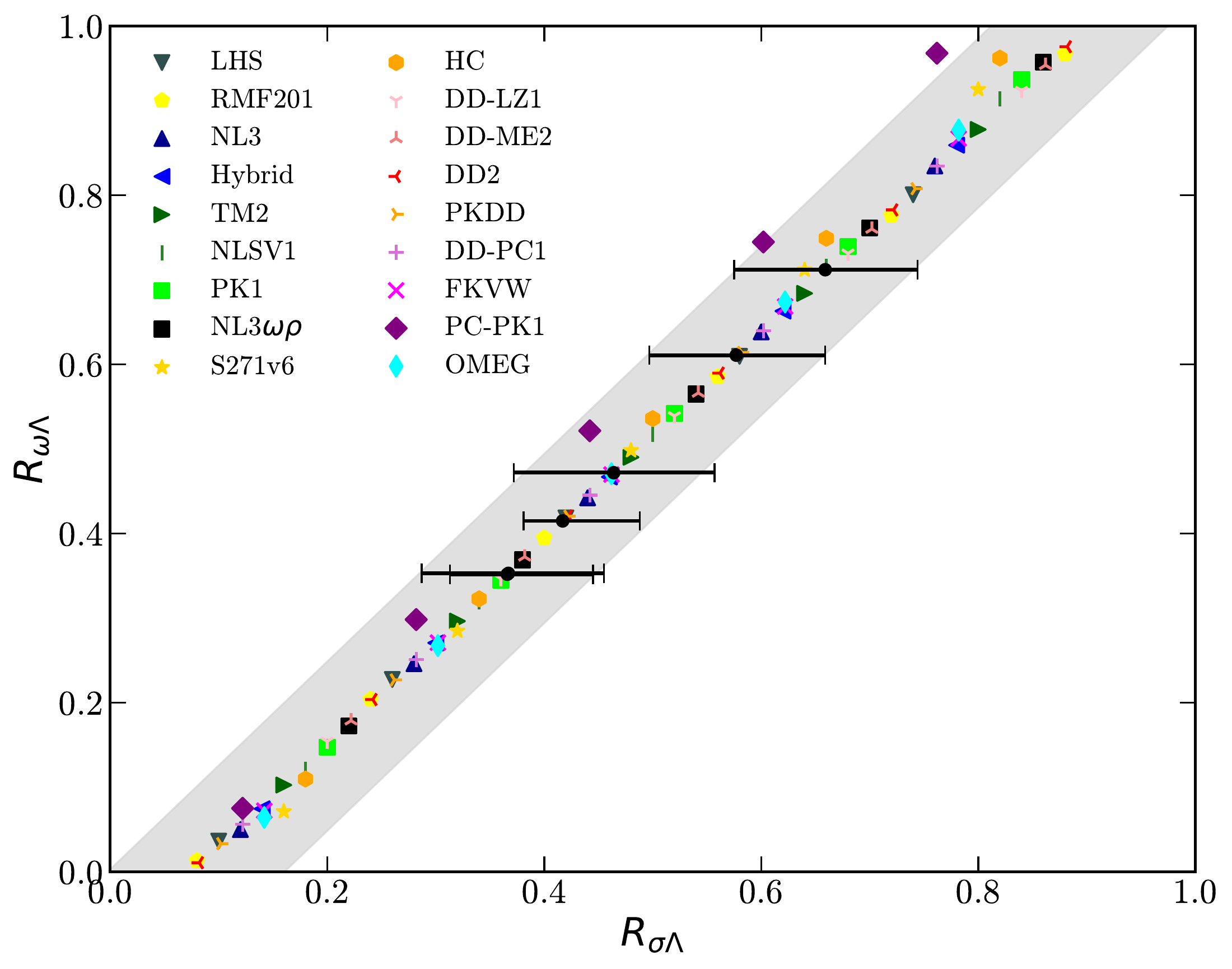} 	\vspace{-0.3cm}
\caption{Linear correlation between $R_{\sigma\Lambda}$ and $R_{\omega\Lambda}$ for the 18 RMF models (LHS, RMF201, NL3, Hybrid, TM2, NLSV1, PK1, NL3$\omega\rho$, S271v6, HC, DD-LZ1, DD-ME2, DD2, PKDD, DD-PC1, PK1, FKVW, PC-PK1 and OMEG) used in this work, obtained from Eq.~(\ref{eq:RMF}) or Eq.~(\ref{eq:PCRMF}). The potential depths are fixed at $U_N=-70\,{\rm MeV}$, $U_\Lambda=-30\,{\rm MeV}$ following~\citet{2021PhRvC.104e4321R}. Also shown are the error of $R_{\sigma\Lambda}$ from fitting the experimental data of $\Lambda$ separation energy for PKDD and DD-ME2, as taken from~\citet{2021PhRvC.104e4321R}. 
The shaded region is the 68\% credible region of our NUCL constraint: $R_{\omega\Lambda}=1.228R_{\sigma\Lambda}-0.097$ with an error of $\sigma_{R_{\sigma\Lambda}}=0.08$ (see above in Sec.~\ref{sec:analysis}).
} 	\vspace{-0.3cm}
\label{fig:linear}
\end{figure}

In Fig.~\ref{fig:linear}  we show the linear correlation between $R_{\sigma\Lambda}$ and $R_{\omega\Lambda}$ for all 18 RMF models used in this work, where the potential depths are set as $U_N=-70\mev$ and $U_\Lambda=-30\mev$ following \citet{2021PhRvC.104e4321R}. 
It is seen that the results of different RMF models locate inside the adopted NUCL constraining region, which justifies the adopted linear relation from hypernulei calculations can be generally applied to our chosen set of RMF models for the study of stellar properties.

\section{Changing the statistical error $\sigma_{R_{\sigma\Lambda}}$ for the NUCL constraint} \label{A2}

In~\citet{2021PhRvC.104e4321R}, the values of $R_{\sigma\Lambda}$ and $R_{\omega\Lambda}$ have been obtained by reproducing the experimental $\Lambda$ separation energies of several single $\Lambda$ hypernuclei. 
Due to the strong correlation between $R_{\sigma\Lambda}$ and $R_{\omega\Lambda}$, \citet{2021PhRvC.104e4321R} only needs to evaluate the errors of the independen parameter $R_{\sigma\Lambda}$.
As seen in Fig.~\ref{fig:linear}, the statistical errors $\sigma_{R_{\sigma\Lambda}}$ reported in \citet{2021PhRvC.104e4321R} for PKDD and DD-ME2 vary slightly from $\sim0.04$ to $0.10$.
In Sec.~\ref{sec:analysis} and Sec.~\ref{sec:result}, the results with $\sigma_{R_{\sigma\Lambda}}=0.08$ are discussed for different employed RMF models.
Here we reperform the analysis with two other values $\sigma_{R_{\sigma\Lambda}}=0.06$ and $0.10$ to examine the sensitivity of our results on $\sigma_{R_{\sigma\Lambda}}$.

\begin{figure}
\vspace{-0.2cm}
\centering 
\includegraphics[width = 0.56\linewidth]{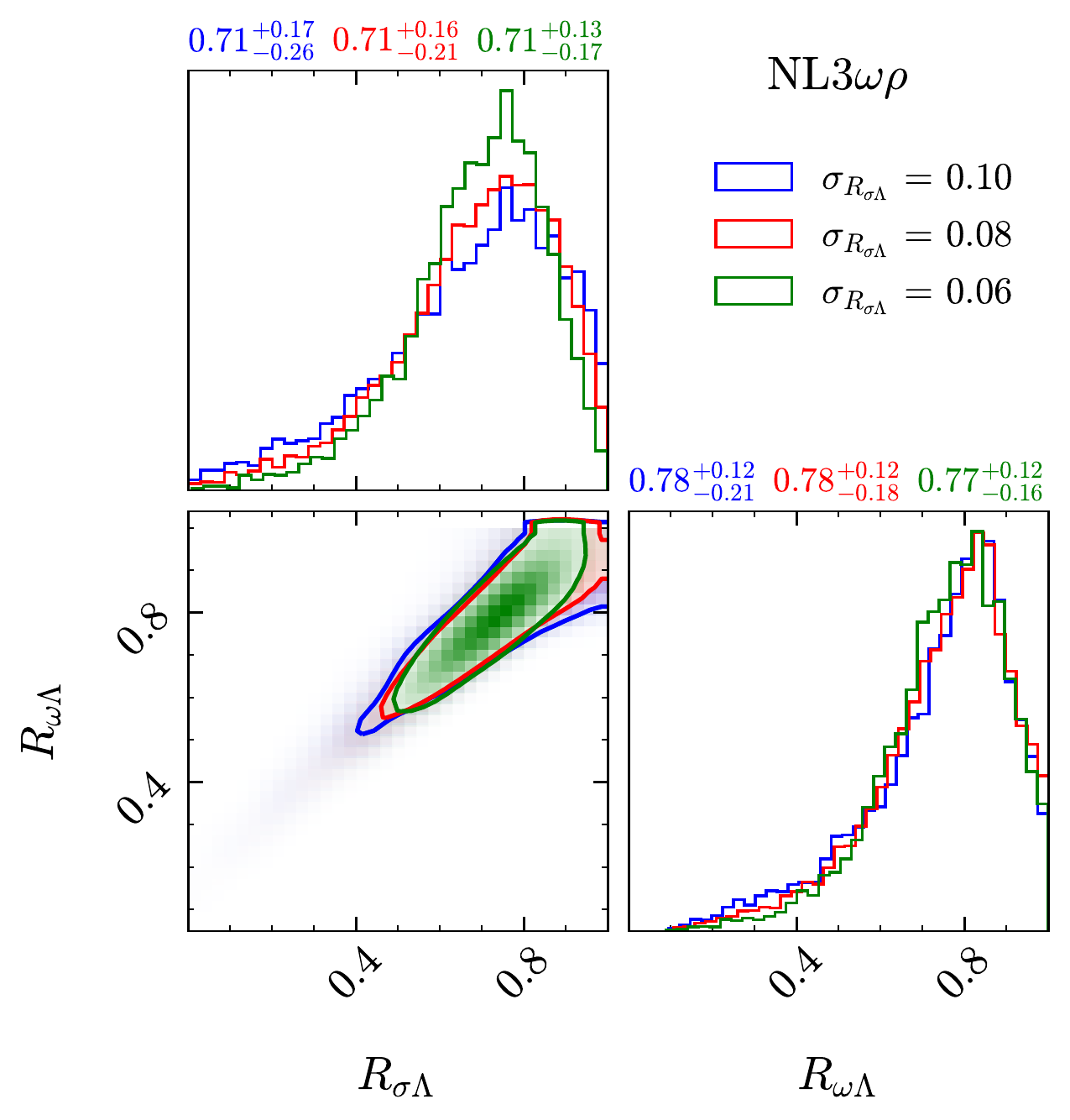} \vspace{-0.3cm}
\caption{Posterior distributions of $R_{\sigma\Lambda}$ and $R_{\omega\Lambda}$ conditioned by different $\sigma_{R_{\sigma\Lambda}}$ values for the representative NL3$\omega\rho$ EOS from the +NICER+GW170817+NUCL joint analysis explained in Sec.~\ref{sec:analysis}.
} 	\vspace{-0.2cm}
\label{fig:3sigma}
\end{figure}

For the representative NL3$\omega\rho$ EOS, in Fig.~\ref{fig:3sigma} we show the posterior distributions of the ratios $R_{\sigma\Lambda}$ and $R_{\omega\Lambda}$, conditioned by different choices of the value of $\sigma_{R_{\sigma\Lambda}}$ as jointly constrained by the +NICER+GW170817+NUCL analysis; We also tabulate the results of the ratios and various hyperon star properties in Table~\ref{tab:sigma}. We see that both the coupling ratios and the hyperon star properties do not rely sensitively on the choice of the statistical error $\sigma_{R_{\sigma\Lambda}}$.

\begin{table*}
\vspace{-0.3cm}
\footnotesize
	\centering
	\caption{Most probable intervals of the ratios $R_{\sigma\Lambda}$ and $R_{\omega\Lambda}$ and various hyperon star properties, at $68\%$ credible level, conditioned by different choices of the value of $\sigma_{R_{\sigma\Lambda}}$ for the representative NL3$\omega\rho$ EOS from the +NICER+GW170817+NUCL joint analysis explained in Sec.~\ref{sec:analysis}. $M_{\rm max}$ is the maximum mass and $\rho_c/\rho_0$ is the corresponding central density scaled by the saturation density. $R_{2.0}$ is the radius of $2.0\Msun$ stars. $R_{1.4}$ and $\Lambda_{1.4}$ are the radius and tidal deformability of $1.4\Msun$ stars, respectively. }
	\label{tab:sigma}
	\vspace{-0.3cm}
\renewcommand\arraystretch{1.4}
  \begin{ruledtabular}
	\begin{tabular}{ccccccccc} 
			  &$R_{\sigma\Lambda}$&$R_{\omega\Lambda}$&$M_{\rm max}/{\rm M_\odot}$ &$\rho_c/\rho_0$ &$R_{2.0}/{\rm km}$ &$R_{1.4}/{\rm km}$ & $\Lambda_{1.4}$     \\
			\hline
			$\sigma_{R_{\sigma\Lambda}}=0.10$
			&$0.707^{+0.174}_{-0.261}$&$0.777^{+0.116}_{-0.209}$&$2.180^{+0.062}_{-0.228}$&$4.865^{+0.055}_{-0.565}$&$13.980^{+0.088}_{-1.533}$&$13.769^{+0.000}_{-1.086}$&$940.165^{+0.000}_{-448.040}$  \\
			\hline
			{$\sigma_{R_{\sigma\Lambda}}=0.08$}
			&$0.712^{+0.157}_{-0.215}$&$0.778^{+0.121}_{-0.183}$&$2.176^{+0.085}_{-0.202}$&$4.846^{+0.046}_{-0.501}$&$13.968^{+0.096}_{-1.512}$&$13.769^{+0.000}_{-1.084}$&$940.165^{+0.000}_{-443.756}$  \\
			\hline
            {$\sigma_{R_{\sigma\Lambda}}=0.06$}
			&$0.710^{+0.135}_{-0.173}$&$0.772^{+0.120}_{-0.159}$&$2.168^{+0.075}_{-0.186}$&$4.966^{+0.032}_{-0.475}$&$13.950^{+0.075}_{-1.423}$&$13.769^{+0.000}_{-1.064}$&$940.165^{+0.000}_{-430.676}$  \\
	\end{tabular}
\end{ruledtabular}
\end{table*}

\section*{Acknowledgements}
The work is supported by National SKA Program of China (No.~2020SKA0120300), the National Natural Science Foundation of China (Grants No.~12273028 and No.~11875152) and the Youth Innovation Fund of Xiamen (No. 3502Z20206061).

\software{Bilby \citep[version 0.5.5, ascl:1901.011, \url{https://git.ligo.org/lscsoft/bilby/}]{2019ascl.soft01011A}, PyMultiNest \citep[version 2.6, ascl:1606.005, \url{https://github.com/JohannesBuchner/PyMultiNest}]{2016ascl.soft06005B}, Toast \citep[\url{https://git.ligo.org/francisco.hernandez/toast}]{2020MNRAS.499.5972H}, Corner \citep[][\url{https://github.com/dfm/corner.py}]{2016JOSS....1...24F}.}

\end{document}